\documentclass[aps,prc,twocolumn,showpacs,preprintnumbers,
               nofootinbib,float,superscriptaddress,longbibliography]{revtex4-1}
\usepackage{graphicx, fancybox}
\usepackage{amsmath,amssymb}
\usepackage{color}
\usepackage[dvipsnames]{xcolor}
\usepackage[colorlinks=true, pdfstartview=FitV, linkcolor=Orange, citecolor=Cyan, urlcolor=blue]{hyperref}
\usepackage{soul}
\usepackage{array}
\usepackage{url}
\usepackage[utf8]{inputenc}
\usepackage[normalem]{ulem}
\usepackage{comment}

\newcommand{\be}{\begin{equation}}
\newcommand{\ee}{\end{equation}}
\newcommand{\ba}{\begin{eqnarray}}
\newcommand{\ea}{\end{eqnarray}}
\newcommand{\bd}{\begin{displaymath}}
\newcommand{\ed}{\end{displaymath}}

\renewcommand{\vec}[1]{\mbox{\boldmath$#1$}}

\begin{document}

\title{Stochastic hydrodynamics and hydro-kinetics: Similarities and differences}
\author{Aritra De}
\affiliation{School of Physics \& Astronomy, University of Minnesota, Minneapolis, MN 55455, USA}
\author{Chun Shen}
\affiliation{Department of Physics and Astronomy, Wayne State University, Detroit, MI 48201, USA}
\affiliation{RIKEN BNL Research Center, Brookhaven National Laboratory, Upton, NY 11973, USA}
\author{Joseph I. Kapusta}
\affiliation{School of Physics \& Astronomy, University of Minnesota, Minneapolis, MN 55455, USA}

\date{\today}

\begin{abstract}
The hydro-kinetic formalism has been used as a complementary approach to solving the Stochastic Differential Equations (SDE) corresponding to noisy hydrodynamics. The hydro-kinetic formalism consists of a deterministic set of relaxation type equations that tracks the evolution of 2-point correlation functions of stochastic hydrodynamic quantities. Hence they are comparatively easier to solve than the SDEs, which are computationally intensive and need to deal with arbitrarily large gradients. This work compares the two approaches for the propagation and diffusion of conserved charge fluctuations in the Bjorken hydrodynamic model. For white noise, the two approaches agree.  For colored Catteneo noise, which is causal, the two approaches diverge.  This is because white noise only induces two-point correlations, while Catteneo noise also induces higher-order correlations. This difference is quantified from the effects of causal evolution and influence from higher-order correlations induced by the Catteneo noise. 
\end{abstract}

\maketitle

\section{Introduction}
Experimental facilities such as the Relativistic Heavy-Ion Collider (RHIC) and the Large Hadron Collider (LHC) collide nuclei at high energies to study and quantify properties of the quark-gluon plasma.
Relativistic hydrodynamics provides a macroscopic and quantitative description of the fireball evolution in heavy-ion collisions~\cite{Heinz:2013th, Gale:2013da, Shen:2020mgh}. The applicability of hydrodynamics is justified if the mean free paths of the particles are small compared to the distances over which thermodynamic quantities vary. It has been successfully used to constrain various transport properties of the quark-gluon plasma such as specific shear and bulk viscosity and thermal conductivity \cite{Pratt:2015zsa, Bernhard:2016tnd, Bernhard:2019bmu, JETSCAPE:2020shq, JETSCAPE:2020mzn, Nijs:2020roc, Nijs:2020ors}.
Hybrid frameworks that combine hydrodynamics with a hadronic transport model can reproduce various flow measurements with successful predictive power \cite{Song:2010mg, Shen:2014vra, Putschke:2019yrg, Schenke:2020mbo}.

Relativistic hydrodynamics involves differential equations for the energy-momentum tensor and the charge currents.
Ripples caused by stochastic thermal fluctuations will give corrections to macroscopic variables of hydrodynamics \cite{Young:2014pka}. The fluctuation-dissipation theorem \cite{Kapusta:2011gt} relates dissipative properties to the amplitude of hydrodynamic fluctuations. Hence, study of fluctuations can be leveraged to study quark-gluon plasma transport properties \cite{Pratt:2016lol, Pratt:2018ebf}. Specifically, fluctuations are known to become important near critical points \cite{Plumberg:2017tvu, Bzdak:2019pkr, Bluhm:2020mpc, An:2021wof}. This has led to studies that investigate fluctuations in conserved quantities, such as electric charge, baryon number, and strangeness on an event-by-event basis \cite{Kapusta:2017hfi,  Pratt:2017oyf, Pratt:2019pnd, De:2020yyx}. Recently, there have also been studies of dynamic critical phenomena near a QCD critical point \cite{Nahrgang:2020yxm, Rajagopal:2019xwg, Du:2020bxp, An:2021wof, Du:2021zqz}. Non-Gaussianities (higher-order cumulants) also become important near critical points as they are more sensitive to large correlation lengths \cite{Bzdak:2019pkr}.

It is important to systematically study fluctuations in heavy-ion collision systems. Adding noise to hydrodynamic equations transforms the hydrodynamical equations from partial differential equations to stochastic differential equations. 
Historically, stochastic differential equations have been notorious to deal with \cite{Young:2014pka, Murase:2015oie, Singh:2018dpk, Sakai:2020pjw, Sakai:2021pev}. Broadly, there have been two methods to study the effect of fluctuations. The first is the direct method of studying stochastic differential equations through analytical and numerical techniques \cite{Kapusta:2017hfi, Young:2014pka, Singh:2018dpk, Sakai:2020pjw, Pratt:2017oyf, Pratt:2019pnd, De:2020yyx}. One can simulate stochastic differential equations like usual partial differential equations over millions of different realizations of the noise. Statistical properties of those solutions are studied to understand the underlying physical process. Alternatively, some authors have suggested the use of a set of deterministic kinetic equations to capture the behavior of time evolution of $n$-point correlations of hydrodynamic variables \cite{Akamatsu:2016llw, Stephanov:2017ghc, Akamatsu:2018vjr, An:2019osr, Martinez:2019bsn, An:2019csj,Martinez:2017jjf}. This approach has come to be known as the hydro-kinetic approach. The method relies on the truncation of higher moments from noisy hydrodynamic calculations and holds the advantage over direct numerical studies of SDEs by being computationally less intensive.  

It seems imperative that a direct comparison between the hydro-kinetic approach and the stochastic hydro formalism needs to be made which is lacking in the current literature. In this paper, we compare the two approaches for a system with just one conserved charge in an expanding fluid background, namely the Bjorken system \cite{Bjorken:1982qr}. At the top RHIC and LHC energies, the charge current decouples from the energy-momentum conservation because the net charge is small and therefore can be studied separately. The system with a single conserved charge is simple enough to allow for analytical calculations. Additionally, the conserved baryon density is believed to be an order parameter for a possible critical point in the QCD phase diagram \cite{Bzdak:2019pkr}. The usual diffusion equation has an infinite signal propagation speed. To make it finite, we also study the Catteneo equation and which introduces so-called colored noise in the hydrodynamical equations \cite{Kapusta:2014dja, Kapusta:2014aza}. In the presence of colored noise, the hydrodynamics become causal. We show the comparison of hydro-kinetics to causal hydrodynamics as well.

The paper is organized as follows. We give a concise description of the dynamical evolution of one conserved charge in a 1+1D Bjorken system in Sec.~\ref{Sec:3DDiffusion}. In Sec.~\ref{Sec:3DKinetic}, we derive the hydro-kinetic equations and compare them with the stochastic calculation in the white noise limit. We extend the comparison for the colored noise case to hydro-kinetics in Sec.~\ref{Sec:3DComp}.
Sec.~\ref{Sec:Conclusion} summarizes our concluding remarks. Complementary calculations without transverse dynamics are provided in the Appendix~\ref{Sec:1DDiffusion}.

\section{Stochastic diffusion in boost-invariant hydrodynamics}\label{Sec:3DDiffusion}

The assumption of longitudinal boost invariance implies that the initial conditions for local variables are functions of the proper time $\tau \equiv \sqrt{t^2 - z^2}$ and the transverse coordinates $x$ and $y$. They do not depend on spacetime rapidity $\xi \equiv \tanh^{-1}\left(z/t \right)$.  We define fluid velocity $u^{\mu}$ to be the velocity of the energy-momentum current, known as the Landau-Lifshitz frame,
\be
T^{\mu\nu} u_\nu = e u^\mu.
\ee
We consider a single species of conserved charge in the fluid. This charge current takes the form 
\begin{equation}
J^{\mu}  =  n_Q u^{\mu} + \Delta J^{\mu} + I^{\mu}, 
\end{equation}
where $n_Q$ is the charge density in the local rest frame of the fluid cell, $\Delta J^{\mu}$ is the diffusion current, and $I^{\mu}$ is the noise current. The fluctuation-dissipation theorem dictates that dissipation is always accompanied by stochastic noises for any equilibrium system. In the conventional, lowest order in derivatives diffusion equation, $\Delta J^{\mu}$ takes the form
\begin{equation}
\Delta J^{\mu} =  D_Q \nabla^{\mu}n_Q = \sigma_Q \, T \, \nabla^{\mu} \left(\frac{\mu_Q}{T}\right),
\end{equation}
where $\mu_Q$ is the charge chemical potential, $\sigma_Q$ is the charge conductivity, and $\nabla^{\mu}$ is the transverse derivative
\begin{equation}
\nabla^{\mu} \equiv \Delta^{\mu\nu} \partial_\nu = \partial^{\mu} - u^{\mu}(u\cdot \partial)
\end{equation}
with the projection operator $\Delta^{\mu\nu} = g^{\mu\nu} - u^\mu u^\nu$.
One can verify that the charge current conservation equation $\partial_\mu J^\mu = 0$ in the local rest frame where $u^\mu = (1,\vec{0})$ follows the usual diffusion equation
\begin{equation}
\left( \frac{\partial }{\partial t} -D_Q \nabla^2  \right) n_Q =0  \,.
\end{equation}
The diffusion constant $D_Q$ and charge conductivity are related by the Einstein relation $D_Q = \sigma_Q/\chi_Q$, where $\chi_Q$ is the electric charge susceptibility defined by
\begin{equation}
\chi_Q = \frac{\partial n_Q(T,\mu_Q)}{\partial \mu_Q} \,.
\end{equation}
Following Refs.~\cite{Kapusta:2017hfi, De:2020yyx}, we assume $\chi_Q \propto T^2$.
The amplitudes for dissipation and stochastic noise are related by the fluctuation-dissipation theorem which is justified because the fluid description assumes the system is close to local equilibrium. In other words, modes disturbed by $I^{\mu}$ are equilibrated by the viscous term $\Delta J^{\mu}$. The 1-point function of the noise $\langle I^\mu \rangle$ vanishes while the 2-point functions are determined by the fluctuation-dissipation theorem. Assuming white noise
\begin{equation}
\langle I^{\mu}(x)\rangle =0, \quad \langle I^{\mu}(x_1)I^{\nu}(x_2) \rangle = 2 \sigma_Q T \, \Delta^{\mu\nu} \delta(x_1-x_2).
\end{equation}
Such a noise function whose 2-point correlation is proportional to the Dirac $\delta$-function in both space and time is referred to as white noise.

We adopt Milne coordinates which are proper time and space-time rapidity,
\begin{eqnarray}
\begin{aligned}
t &=& \tau \cosh \xi \qquad z &= \tau \sinh \xi \\
\tau &=& \sqrt{t^2-z^2} \qquad \xi &= \tanh^{-1}\left( \frac{z}{t} \right). \\
\end{aligned}
\end{eqnarray}
The transverse coordinates $x$ and $y$ are the same as in Cartesian coordinates. The non-zero components of the background flow velocity are
\begin{equation}
u^0 = \cosh \xi \qquad u^z = \sinh \xi \,.
\end{equation}
The transverse derivatives in the $t$ and $z$ directions are
\begin{eqnarray}
\nabla^0 = -\frac{\sinh \xi}{\tau} \frac{\partial }{\partial \xi} \qquad \nabla^3 = -\frac{\cosh \xi}{\tau} \frac{\partial }{\partial \xi}
\end{eqnarray}
with $\, u\cdot \partial = \partial/\partial \tau$. The entropy density $s$ can be factored out from the noise current $I^{\mu}$ so as to capture the fluctuating part in the random dimensionless scalars $f$, $g_1$ and $g_2$. 
\begin{eqnarray}
I^t &=& s(\tau)\; f(\xi,\tau,x,y)\sinh \xi \nonumber \\
I^z &=& s(\tau) \;f(\xi,\tau,x,y)\cosh \xi \nonumber \\
I^x &=& s(\tau) \; g_1(\xi,\tau,x,y) \nonumber \\
I^y &=& s(\tau) \; g_2(\xi,\tau,x,y) 
\end{eqnarray}
Here $f,g_1,g_2$ are uncorrelated random scalar functions with the following 2-point correlation in Cartesian coordinates
\begin{eqnarray}
\langle f(x_1)f(x_2) \rangle &=& \langle g_1(x_1)g_1(x_2) \rangle = \langle g_2(x_1)g_2(x_2) \rangle \nonumber \\
&=& \frac{2\sigma_Q T}{s^2}\delta^4(x_1-x_2)\; \;\;
\label{corr_1}
\end{eqnarray}
and \begin{equation}
\langle f(x_1)g_1(x_2) \rangle = \langle g_2(x_1)g_1(x_2) \rangle = \langle f(x_1)g_2(x_2) \rangle= 0.
\label{uncorrelated}
\end{equation}
The last equation denotes that the random functions $f$, $g_1$ and $g_2$ are uncorrelated with each other.

The background conservation equations for the charge density and entropy density are 
\begin{eqnarray}
\frac{d s}{d \tau} + \frac{s}{\tau} = 0 \;\; &\Rightarrow& \;\; s(\tau) = \frac{s_i \tau_i}{\tau} \\
\frac{d n_Q}{d \tau} + \frac{n_Q}{\tau} = 0 \;\; &\Rightarrow& \;\; n_Q(\tau) = \frac{n_i \tau_i}{\tau} \,.
\end{eqnarray}
The $s_i$ and $n_i$ are the entropy and charge densities at some initial time $\tau_i$. We take the initial charge density $n_i$ to be zero, hence the average charge density for subsequent times is zero as well. Please note that because the stochastic fluctuations are small perturbations compared to the background quantities, we neglect their feedback corrections to the background fluid velocity in this study.

We now have all the pieces of the charge current $J^{\mu}$ to write the charge current conservation equation $\partial_{\mu}J^{\mu} = 0$.  It is convenient to define the variable $X = \tau \delta n_Q$. The charge conservation equation then becomes
\begin{eqnarray}
\frac{1}{\tau}\frac{\partial X}{\partial \tau} - \frac{D_Q}{\tau} \left[ \nabla_{\perp}^2 + \frac{1}{\tau^2}\frac{\partial^2 }{\partial \xi^2} \right]X = -\frac{s}{\tau} \frac{\partial f}{\partial \xi} - s \frac{\partial g_1}{\partial x} - s\frac{\partial g_2}{\partial y} \nonumber \\
\label{charge}
\end{eqnarray}
where $\nabla^2_\perp = \partial^2/\partial x^2 + \partial^2/\partial y^2$.
We do a Fourier transform
\begin{equation}
X(\xi, x, y, \tau) = \int \frac{dk_\xi}{2\pi} \frac{d^2 \mathbf{k_{\perp}}}{2\pi}
e^{i(k_\xi \xi + \mathbf{k_{\perp}.x_{\perp}})} \; \tilde{X} (k_\xi,\mathbf{k_{\perp}},\tau).
\end{equation}
In Fourier space the charge conservation equation becomes
\begin{equation}
\frac{\partial \tilde{X}}{\partial \tau} + D_Q\mathbf{k}_{\perp}^2\tilde{X} + \frac{D_Qk_\xi^2}{\tau^2}\tilde{X} = - is \left[k_\xi \tilde{f} + \tau k_x \tilde{g_1}+ \tau k_y\tilde{g_2}  \right].
\label{3D_charge_conservation}
\end{equation}
The Green's function of the homogeneous part of the above equation is
\begin{equation}
\tilde{G}(k_\xi,\mathbf{k_{\perp}},\tau_f,\tau_i) = e^{-D_Q\mathbf{k}_{\perp}^2(\tau_f-\tau_i)}
e^{-D_Q k_\xi^2 \left( \frac{1}{\tau_i}-\frac{1}{\tau_f}\right)}.
\end{equation}
Then
\begin{eqnarray}
\tilde{X}(k_\xi, k_x, k_y, \tau_f) &=& - i \int^{\tau_f}_{\tau_i} d\tau s 
\left[ k_\xi \tilde{f} + \tau k_x\tilde{g_1} + \tau k_y\tilde{g_2} \right]  \nonumber \\
& \times& e^{-D_Q\mathbf{k}_{\perp}^2(\tau_f-\tau_i)}
e^{- D_Qk_\xi^2\left( \frac{1}{\tau_i}-\frac{1}{\tau_f}\right)}.
\end{eqnarray}
The 2-point correlation function can now be calculated as
\begin{eqnarray}
&& \langle\delta\tilde{n}_Q(k_\xi,\mathbf{k_{\perp}},\tau_f) \, \delta \tilde{n}_Q(-k_\xi',-\mathbf{k'_{\perp}},\tau_f) \rangle \nonumber \\
&=& \frac{4\pi D_Q \chi_f T_f}{\tau_f} \int d\tau' \left[ \frac{k_\xi^2}{\tau'^2} + k_x^2 + k_y^2 \right]\hspace{2cm}\nonumber \\ 
& \times & e^{-2D_Q\mathbf{k}_{\perp}^2(\tau_f-\tau')} 
e^{-2D_Q k_\xi^2\left( \frac{1}{\tau'}-\frac{1}{\tau_f}\right)}
\label{exacttransverse}
\end{eqnarray}
The above integral can be solved analytically only if $k_x = k_y = 0$, see Appendix~\ref{Sec:1DDiffusion}. We will perform a numerical computation of Eq.~\eqref{exacttransverse} in the following section.

\section{The hydro-kinetic equation for the 2-point correlation function}
\label{Sec:3DKinetic}

In this section, we derive and solve the equation of motion for the 2-point function of stochastic fluctuations.  This is referred to as the hydro-kinetic approach.
Let us begin with the derivation of the hydro-kinetic equation by ignoring the transverse directions $x$ and $y$. The noisy charge conservation equation in the absence of transverse noise in $k$-space is
\begin{eqnarray}
\frac{\partial \delta \tilde{n}_Q}{\partial \tau} = -\frac{\delta \tilde{n}_Q}{\tau} - \frac{D_Q k^2}{\tau^2}\delta \tilde{n}_Q - \frac{iks}{\tau}\tilde{f}.
\label{charge_1}
\end{eqnarray}
The noise correlation can be calculated from Eq. \eqref{corr_1} as
\begin{equation}
\langle \tilde{f}(k_1,\tau_1) \tilde{f}(k_2,\tau_2) \rangle = \frac{4\pi \sigma_Q T}{A\tau_1s^2} \delta(\tau_1-\tau_2)\delta(k_1+k_2).
\end{equation}
where $A$ is the transverse area.  The evolution of the 2-point function satisfies
\begin{eqnarray}
\frac{d}{d\tau} \langle \delta \tilde{n}_Q&(k)& \delta \tilde{n}_Q(-k')\rangle \nonumber \\
=&-&\frac{2}{\tau} \langle \delta \tilde{n}_Q(k) \delta \tilde{n}_Q(-k')\rangle \nonumber \\
&-& \frac{D_Q}{\tau^2}(k^2+k'^2)\langle \delta \tilde{n}_Q(k) \delta \tilde{n}_Q(-k')\rangle \nonumber \\
&-&\frac{iks}{\tau}\langle \tilde{f}(k) \delta \tilde{n}_Q(-k')\rangle \nonumber \\
&+& \frac{ik's}{\tau} \langle \delta \tilde{n}_Q(k) \tilde{f}(-k')\rangle.
\label{charge_0}
\end{eqnarray}
We can compute the $\langle \tilde{f} \tilde{n}_Q \rangle$ up to $\mathcal{O}(\delta \tilde{n}_Q^2)$ accuracy using the ideal part of the equation of motion and ignoring the dissipative part.
\begin{eqnarray*}
\delta \tilde{n}_Q =  - \int  \frac{d\tau}{\tau}(iks)\tilde{f} \quad  \Rightarrow \quad \langle \delta \tilde{n}_Q \tilde{f} \rangle = - \int \frac{ d\tau }{\tau} (iks) \langle \tilde{f} \tilde{f} \rangle
\end{eqnarray*}
Then
\begin{eqnarray}
\frac{i k' s}{\tau}\langle \delta \tilde{n}_Q(k)\tilde{f}(-k') \rangle &=& \frac{i k' s}{\tau}\int \frac{ d\tau }{\tau} (-iks )\langle \tilde{f}(k)\tilde{f}(-k') \rangle \nonumber \\
&=& \frac{k k' 4\pi \sigma_Q T}{A \tau^3}\delta(k-k') 
\end{eqnarray}
We now define the charge correlation $N(\tau,k)$ as
\begin{equation}
N(\tau_1,k) \delta(k-k') \equiv \frac{ \langle \delta \tilde{n}_Q(k,\tau_1) \delta \tilde{n}_Q(-k',\tau_1)\rangle}{2\pi}.
\end{equation}
Using this result in Eq. \eqref{charge_0}, we get the following evolution equation for the 2-point correlation function 
\begin{eqnarray}
\frac{d N(\tau,k)}{d\tau}  = -\frac{2D_Q k^2}{\tau^2}  \left( N(\tau,k) - \frac{ \chi_Q T}{A \tau}\right) - \frac{2}{\tau} N(\tau,k) \qquad
\label{kinetic_1}
\end{eqnarray}
Equation~\eqref{kinetic_1} is the hydro-kinetic equation for this situation.  The term $\chi_Q T/A \tau$ is the asymptotic high-$k$ limit. In coordinate space, this term represents self-correlation.

We now generalize the hydro-kinetic equation to 3+1 D. The way we do it is by recognizing that when we include the transverse directions they have dimensions whereas the spacetime rapidity is dimensionless. Therefore the $k$ will be replaced by $\mathbf{K} = (\frac{k_{\xi}}{\tau},k_x,k_y)$. We obtain the evolution equation for the 2-point correlation function in 3D as
\begin{eqnarray}
\frac{dN}{d\tau} = -2D_Q \mathbf{K}^2\left(  N - \frac{\chi_Q T}{A \tau } \right) - \frac{2N}{\tau}.
\label{kinetic3D}
\end{eqnarray}
The form of this evolution equation matches that of Ref.~\cite{Akamatsu:2016llw,Martinez:2017jjf}. 

\begin{figure}[h!]
\centering
\includegraphics[width=0.95\linewidth]{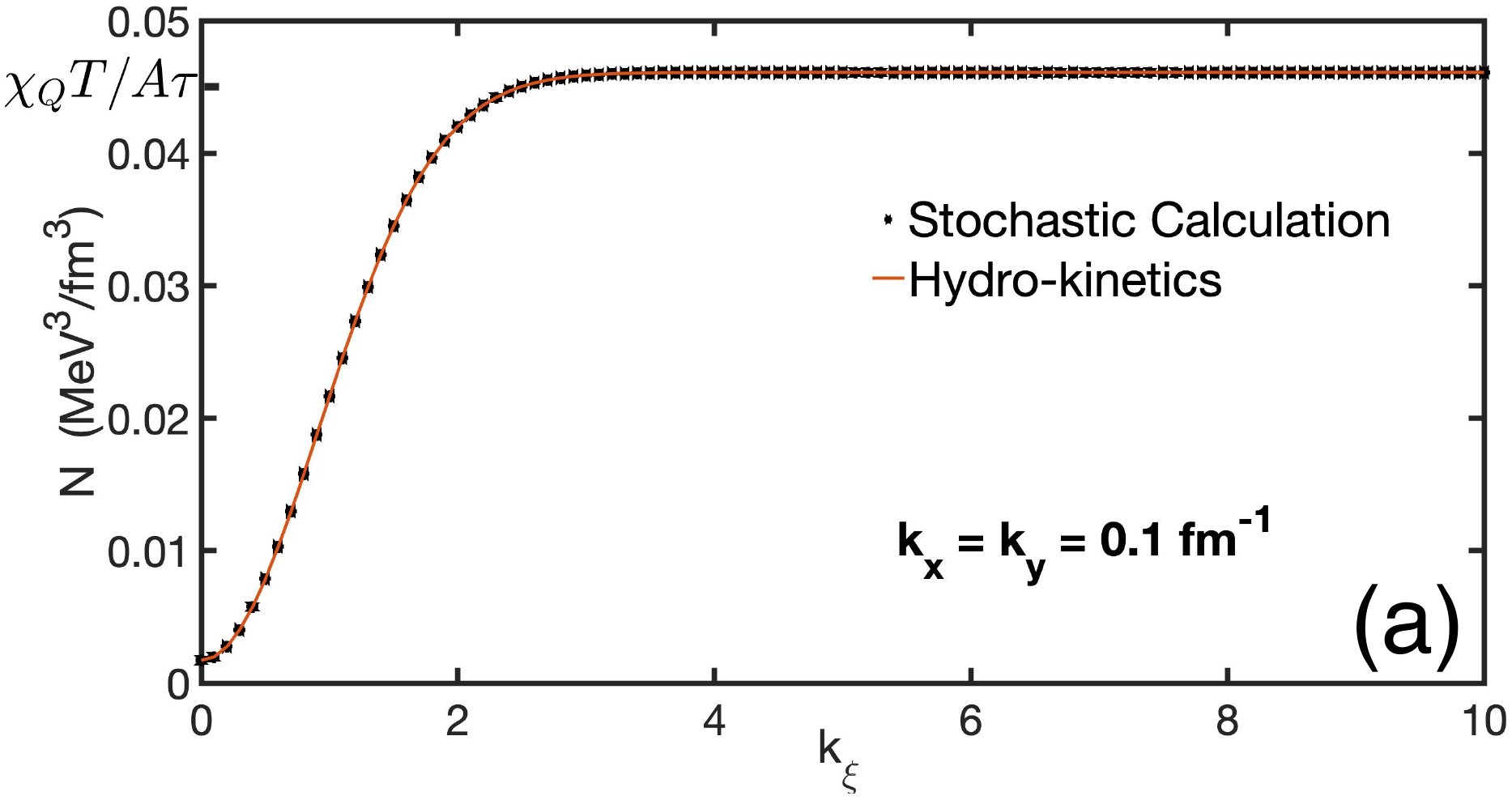}
\includegraphics[width=0.95\linewidth]{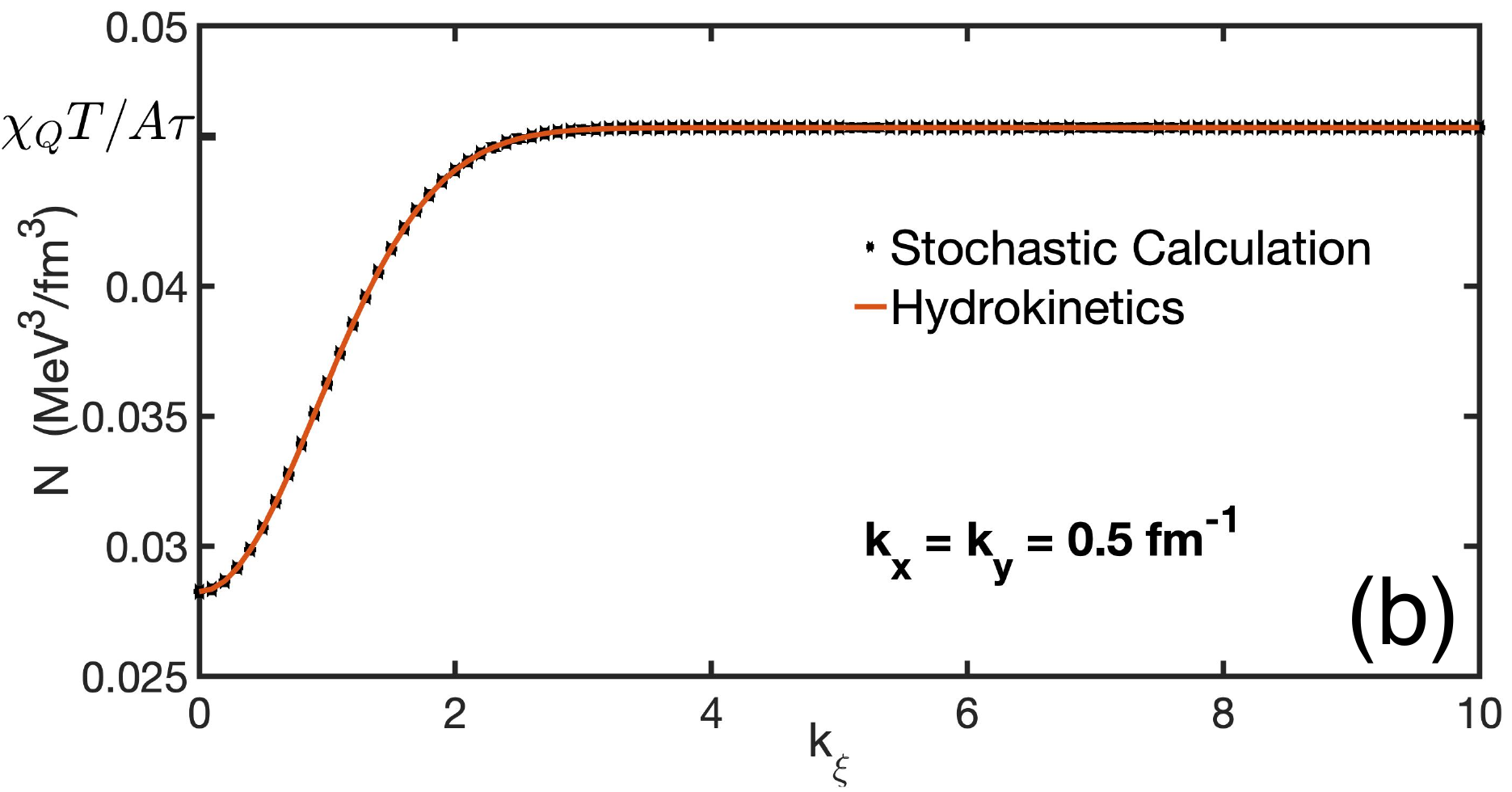}
\caption{(Color online) Comparison of momentum-space 2-point correlation functions for charge density fluctuations from the hydro-kinetics approach and stochastic hydrodynamics. (a) is for $k_x = k_y = 0.1$ fm$^{-1}$ and (b) is for $k_x = k_y = 0.5$ fm$^{-1}$. Both plot are at final time $\tau_f = 6.3$ fm/$c$ with $A = 1$ fm$^2$.}
\label{fig:3Dacausal1}
\end{figure}

We solve the hydro-kinetic equation Eq.~\eqref{kinetic3D} numerically and compare it with the integral Eq.~\eqref{exacttransverse}. Because Eq.~\eqref{kinetic3D} shows the two-point correlation function $N$ only depends on the magnitude of $K$, it is sufficient for us to show the comparisons at two values of $k_x$ and $k_y$ in Fig.~\ref{fig:3Dacausal1}. 
The hydro-kinetic result matches very well with the stochastic solutions. This agreement between the two formalism is expected because the noise functions in Eqs.~\eqref{corr_1} and \eqref{uncorrelated} only introduce 2-point correlations in the white noise limit. There is an emergent length scale at $k_{\xi} = k_*$ from the competition between longitudinal expansion and local equilibration \cite{Akamatsu:2016llw}.
We can equate the viscous dissipation rate to the expansion rate to get $k_*$
\begin{eqnarray}
    2D_Q \left(k_x^2 + k_y^2 + \frac{k_*^2}{\tau_f^2} \right) = \frac{2}{\tau_f}. 
\end{eqnarray}
We set the diffusion constant $D_Q$ to be $0.16$ fm. For $k_x = k_y = 0.1$ fm$^{-1}$ and for the parameters we chose, $k_*$ turns out to be 6.21. It is dimensionless because space-time rapidity is dimensionless. For $k_x = k_y = 0.5$ fm$^{-1}$, $k_*$ is 4.42. One can see in Fig.~\ref{fig:3Dacausal1} that the 2-point correlator has attained the equilibrium value of $\chi_Q T/\tau_f$ for momentum scales $k_{\xi} \gtrsim 0.5 k_*$.
In the following section, we will extend our analysis to causal forms of noise, which are also called colored noise. 

\section{Stochastic noise in causal 3+1D relativistic hydrodynamics}\label{Sec:3DComp}

\begin{figure}[h!]
\centering
\includegraphics[width=0.9\linewidth]{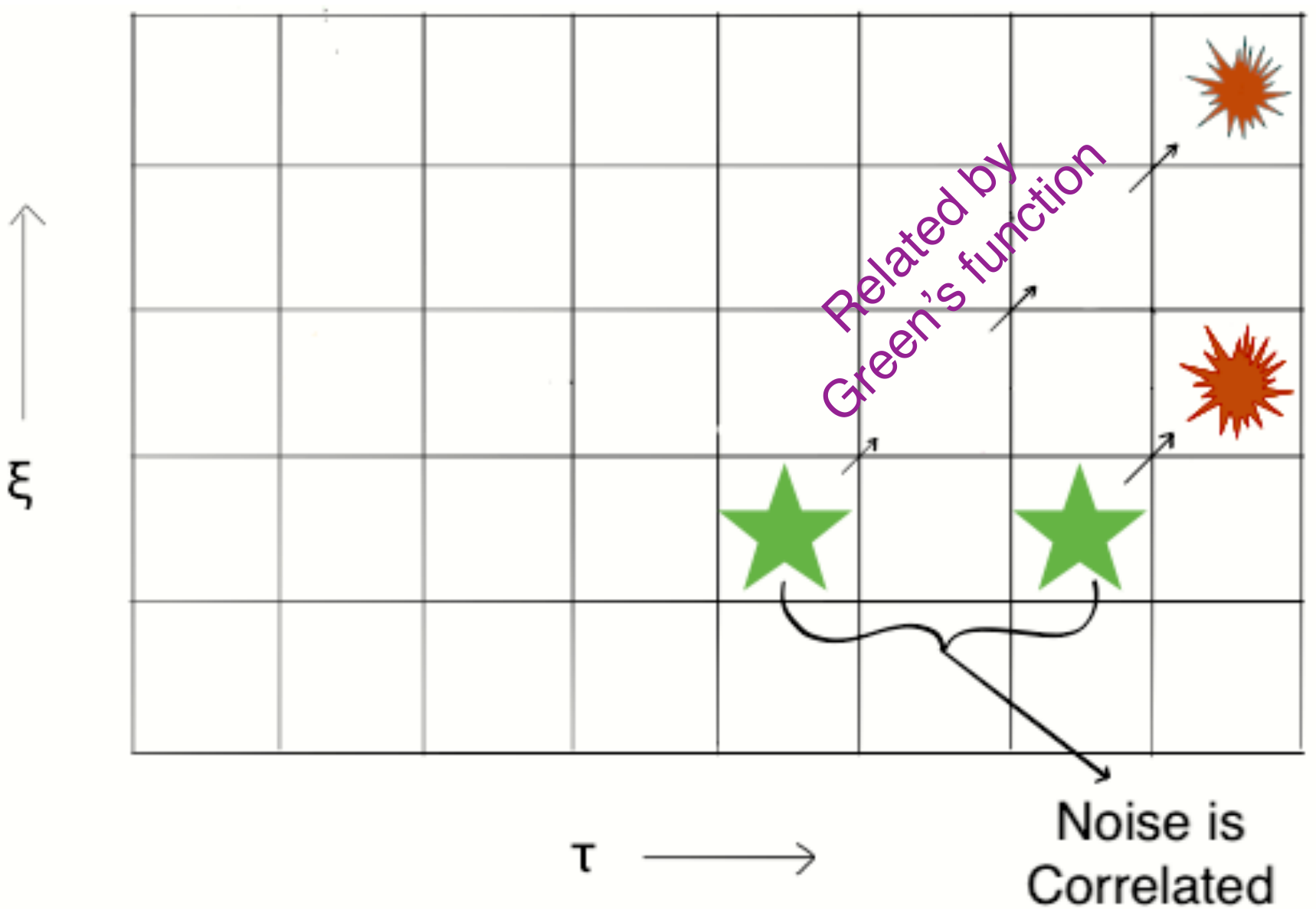}
\caption{(Color online) Illustration of colored noise propagation. The stars denote a noise source and the bursts are the charge fluctuations resulting from noise. Reproduced from Ref.~\cite{De:2020yyx}.}
\label{fig:CorrIllustration}
\end{figure}

The signal propagation speed in the conventional charge diffusion equation is infinite, which violates causality in the relativistic regime. In order to limit signal propagation speeds to less than the speed of light, we introduce a relaxation time $\tau_Q$ in the diffusion equation,
\be
\left( \frac{\partial }{\partial t} -D_Q \nabla^2  + \tau_Q \frac{\partial^2}{\partial t^2} \right) n_Q =0  
\ee
This equation is called the Catteneo equation \cite{cattaneo1958forme} and the noise associated with it is called the Catteneo noise. One can show that high frequency waves travel at a speed of $v_Q = \sqrt{D_Q/\tau_Q}$ \cite{Kapusta:2014aza}. The dissipative current is modified to
\be
\Delta J^{\mu} = D_Q \Delta^{\mu}\left[ \frac{1}{1+\tau_Q(u\cdot \partial )}\right] n_Q. 
\label{viscous_current}
\ee
The fluctuation-dissipation theorem dictates the amplitude of the stochastic noise part of the charge current,
\be
\langle I^{i}(x_1)I^{j}(x_2) \rangle =  \frac{\sigma_Q T}{\tau_Q}  \delta(\vec{x}_1-\vec{ x}_2) \, e^{-|t_1-t_2|/\tau_Q} \, \delta_{ij}.
\label{eq:coloredCorr}
\ee
The Dirac $\delta$-function in time is replaced by an exponential decay function.  In the limit $\tau_Q \rightarrow 0$ this 2-point function becomes the Dirac $\delta$-function for the white noise limit in Eq.~\eqref{corr_1}. Such a noise function with an exponential function replacing the Dirac $\delta$-function in the 2-point correlation is an example of colored noise. If we use the previous parametrization of $I^{\mu}$, the auxiliary function $f$ will have the property
\ba
\langle (\tau_Q \, \partial_{\tau_1} +1)\tilde{f}(k_1,\tau_1) (\tau_Q \, \partial_{\tau_2} +1)\tilde{f}(k_2,\tau_2)\rangle \nonumber \\
 = \frac{4 \pi \sigma_Q (\tau_1) T(\tau_1)}{A \tau_1 s^2(\tau_1)} \, \delta(\tau_1-\tau_2)\delta(k_1 + k_2).
 \label{colored_correlation}
\ea
The same is the case for $g_1$ and $g_2$. The source functions $f, g_1,$ and $g_2$ are still uncorrelated with each other as Eq.~\eqref{uncorrelated} still holds. Noise generated at the same rapidity but at different times is correlated. Figure~\ref{fig:CorrIllustration} illustrates what fluctuation propagation looks like in the presence of the Catteneo noise \cite{De:2020yyx}.
Figure~\ref{fig:CorrIllustration} also illustrates the correlation introduced in the system due to colored noise. For a white noise profile, the two star noise sources will not be correlated and hence the charge fluctuation that has traveled will not have any correlation. But if we switch on colored noise, there will be a correlation between the two bursts.

In the Catteneo equation of motion for the charge current, the viscous current is given by Eq.~\eqref{viscous_current}. The evolution equation for charge fluctuations represented by $X = \tau \delta n_Q$ in momentum space is 
\begin{eqnarray}
\tau_Q \frac{\partial^2 \tilde{X}}{\partial \tau^2} &+& \left( 1+ \frac{2\tau_Q}{\tau} \right) \frac{\partial \tilde{X}}{\partial \tau} + \frac{D_Q k_{\xi}^2}{\tau^2} \tilde{X} +  (k_x^2 + k_y^2) D_Q \tilde{X} \nonumber \\
&+& \frac{2 D_Q \tau_Q (k_x^2 + k_y^2)}{\tau} (1 + \tau_Q \partial_{\tau})^{-1}  \tilde{X} \nonumber \\
&=& -i k_{\xi} s \left[ \tau_Q \frac{\partial \tilde{f}}{\partial \tau} + \left( 1 + \frac{\tau_Q}{\tau} \right) \tilde{f} \right] \nonumber \\
&& - is \bigg[ ( k_x \tilde{g}_1 + k_y \tilde{g}_2) (\tau + 2\tau_Q) \nonumber \\
&& + \tau_Q\tau \left( k_x \frac{\partial \tilde{g}_1}{\partial \tau}  +  k_y \frac{\partial \tilde{g}_2}{\partial \tau} \right) \bigg].
\label{charge_complicated}
\end{eqnarray}
To avoid a third-order differential equation we make the following approximation
\be
    (1+\tau_Q \partial_{\tau})^{-1} \simeq 1-\tau_Q \partial_{\tau}.
    \label{approximation}
\ee
This is a good approximation when $\tau_Q < \tau$. The approximated charge conservation equation is
\begin{eqnarray}
\tau_Q \frac{\partial^2 \tilde{X}}{\partial \tau^2} &+& \left(  1 + 2\frac{\tau_Q}{\tau} \right) \frac{\partial \tilde{X}}{\partial \tau} + \frac{D_Q k_{\xi}^2}{\tau^2} \tilde{X} + \frac{2D_Q \tau_Q (k_x^2 + k_y^2)}{\tau}\tilde{X} \nonumber \\
&-& \frac{2D_Q \tau_Q^2 (k_x^2 + k_y^2)}{\tau} \frac{\partial \tilde{X}}{\partial \tau} + D_Q (k_x^2 + k_y^2) \tilde{X} \qquad \qquad \nonumber \\
&=& -i k_{\xi} s \left[ \tau_Q \frac{\partial \tilde{f}}{\partial \tau} + \left( 1 + \frac{\tau_Q}{\tau} \right) \tilde{f} \right] \nonumber \\
&& - is \bigg[ (k_x \tilde{g}_1 + k_y \tilde{g}_2) (\tau + 2\tau_Q) \nonumber \\
&& + \tau_Q\tau \left( k_x \frac{\partial \tilde{g}_1}{\partial \tau}  +  k_y \frac{\partial \tilde{g}_2}{\partial \tau} \right) \bigg].
\label{equation_approx}
\end{eqnarray}
Eq.~\eqref{equation_approx} is a complex SDE and is challenging to solve analytically. If we remove the dependence on the transverse coordinates it is possible to solve the SDE analytically, as was done in Ref.~\cite{Kapusta:2017hfi}. With the inclusion of transverse coordinates, we will solve the SDE using direct numerical simulations on a discrete lattice. To ensure numerical convergence, we use $d \tau = 0.003$\,fm/$c$. 
Detailed numerical schemes are explained in Appendix~\ref{Sec:stochastic_machinery}. 

\begin{figure}[h!]
\centering
\includegraphics[width=0.95\linewidth]{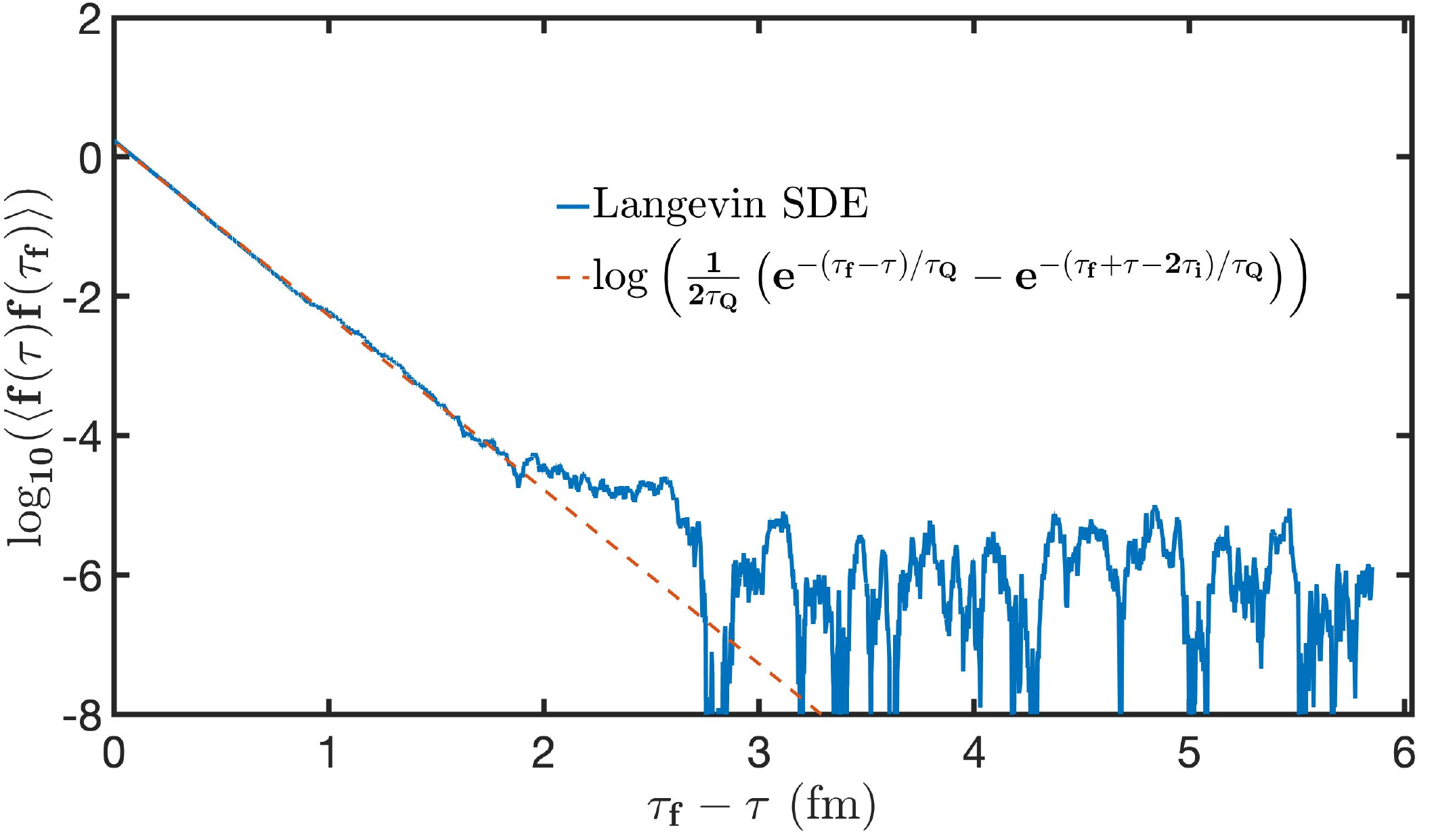}
\caption{(Color online) The time evolution for the variance of fluctuations. Calculations are averaged over one million events with $\tau_f = 6.3$\,fm/$c$, $\tau_i = 0.5$\,fm/$c$,  $\tau_Q = 0.4$\,fm/$c$ and $M = 1$ fm/$c$.}
\label{fig:1DCorrEvo}
\end{figure}

\begin{figure}[h!]
\centering
\includegraphics[width=0.95\linewidth]{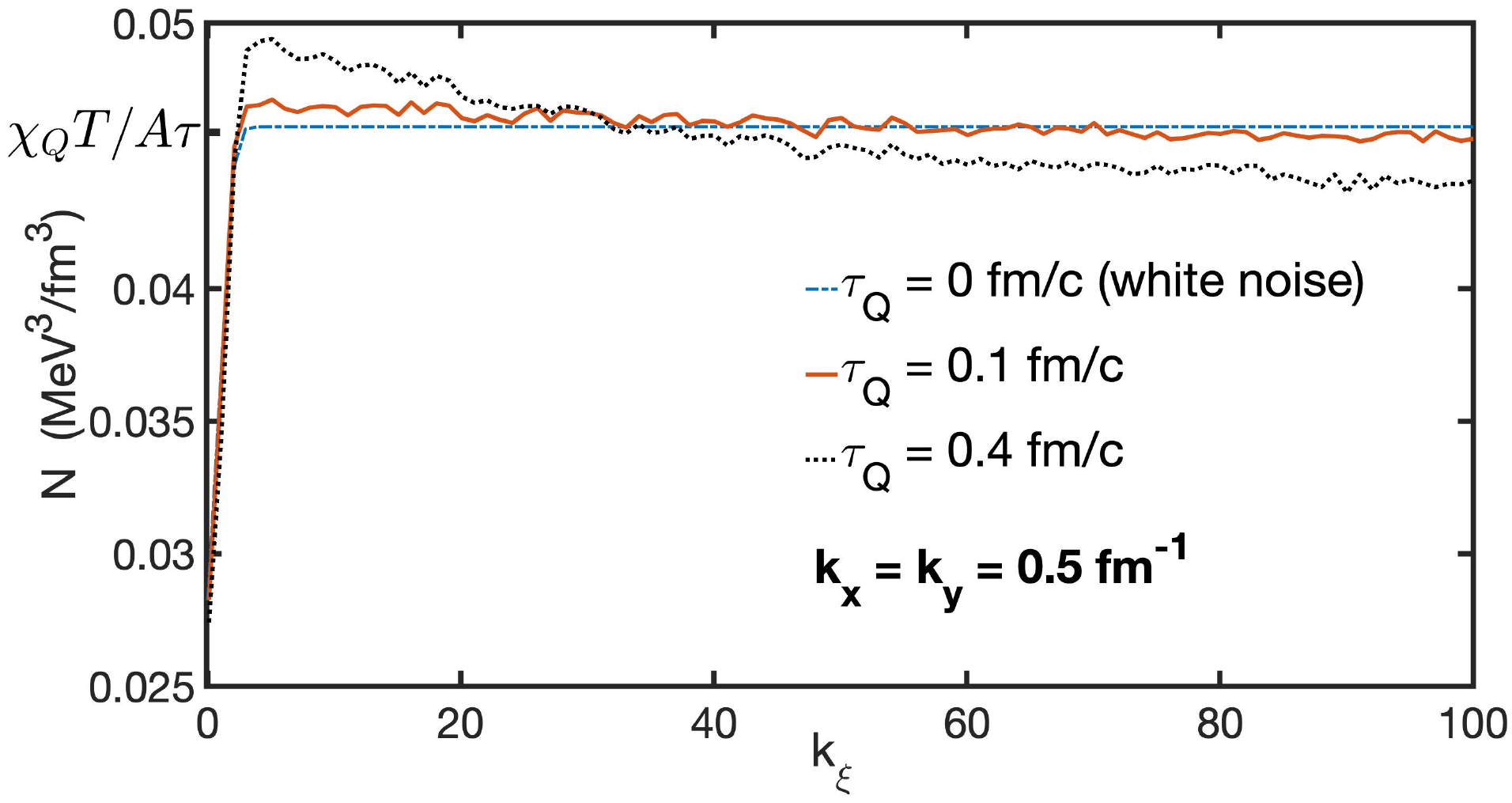}
\caption{(Color online) The 2-point correlation function for charge fluctuations with causal evolution. Results with different $\tau_Q$ values are compared with the white noise limit $\tau_Q = 0$ fm/$c$. The result is for $10^5$ events and is at final time $\tau_f = 6.3$ fm/$c$ with $A = 1$ fm$^2$.}
\label{fig:3Dcausal1}
\end{figure}

To generate a 2-point correlation in the form of Eq.~(\ref{eq:coloredCorr}) for the noise term $f(\tau, \xi)$ we need to solve an additional Langevin equation
\begin{equation}
f + \tau_Q \frac{\partial f}{\partial \tau} = \zeta
\label{eq:LangevinEq}
\end{equation}
where $\zeta$ is a white noise such that 
\be
\langle \zeta(\tau)\rangle  = 0, \qquad \langle \zeta(\tau_1)\zeta(\tau_2)\rangle = M(\tau_1) \delta(\tau_1 - \tau_2).
\ee
Here $M(\tau)$ is the variance of the white noise $\zeta (\tau)$. The analytical solution of the Langevin equation is given by
\begin{equation}
\langle f(\tau_1)f(\tau_2) \rangle   
 = \frac{M}{2 \tau_Q} \left[ e^{|\tau_1-\tau_2|/\tau_Q} - e^{(2\tau_i -\tau_1-\tau_2)/\tau_Q} \right].
\end{equation}
For illustration, Fig.~\ref{fig:1DCorrEvo} uses  $M = 1$ fm/$c$ and shows that the analytical solution and the numerical solution match each other. The noise that we need in this work is given by the 2-point correlation given in Eq.~\eqref{colored_correlation}.
Note that this is the same correlation we need for both $g_1$ and $g_2$.
Once we generate the noise terms we can simulate Eq.~\eqref{equation_approx} for different values of $\tau_Q$ as shown in Fig.~\ref{fig:3Dcausal1}.

\begin{figure}[h!]
\centering
\includegraphics[width=0.95\linewidth]{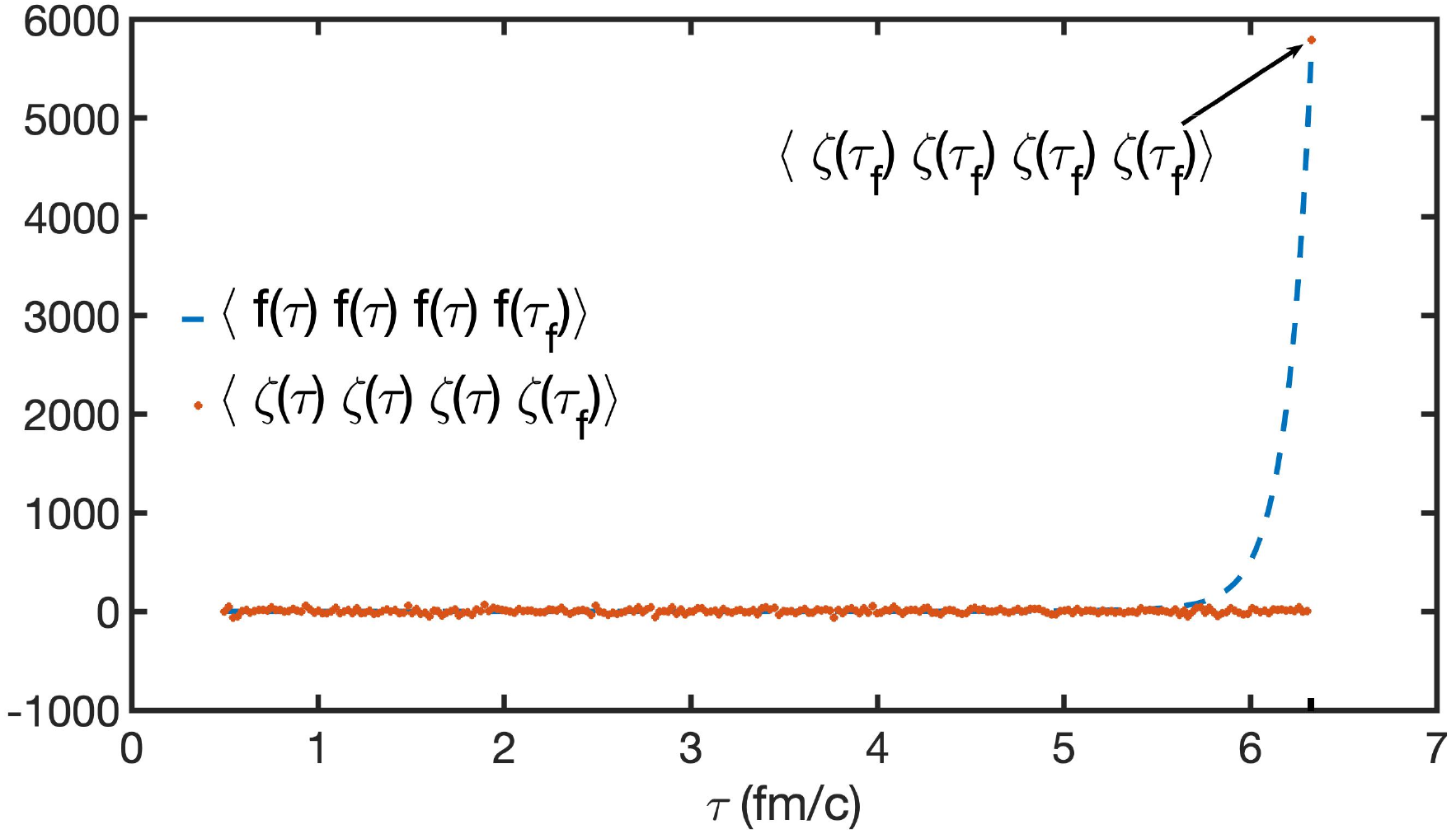}
\caption{(Color online) The 4-point correlation functions for the Catteneo noise $f$ and white noise $\zeta$ with $\tau_Q = 0.4$ fm/$c$.}
\label{fig:4ptcorrelation}
\end{figure}

In the case of the Catteneo noise, the 2-point correlator shows noticeable differences from the results in the white noise limit. It is because the Catteneo noise introduces higher-order correlations that are not tracked by the hydro-kinetic equation, as shown in Fig.~\ref{fig:4ptcorrelation}\footnote{The actual values of the four-point function depends on the lattice spacing used in the simulation. With $\Delta x = 0.013$\,fm, we expect $\langle \zeta \zeta \zeta \zeta \rangle \propto 1/(\Delta x)^2 \sim 6000$.}.
These higher-order correlations are generated from the Langevin equation in Eq.~\eqref{eq:LangevinEq}.  The 4-point correlator of white noise shown in the figure should be strictly zero except when $\tau = \tau_f$.  The numerical simulations reproduce that mathematical statement.  The same 4-point correlator for Catteneo noise need not be zero because of higher-order correlations not present in white noise.  This particular correlator is sufficient to exemplify that fact.  Cumulants could also be used to illustrate that, but calculating them takes about two orders of magnitude more computational time to obtain the same accuracy.  One can see that for Catteneo noise there is no fixed equilibrium value for any finite value of $k_{\xi}$.
For a lower signal propagation speed, meaning larger $\tau_Q$, the magnitude of the 2-point correlation function $N$ is smaller at larger values of $k_{\xi}$. That arises from the nature of the self-correlation term as described in Ref.~\cite{De:2020yyx}.

\begin{figure}[h!]
\centering
\includegraphics[width=0.95\linewidth]{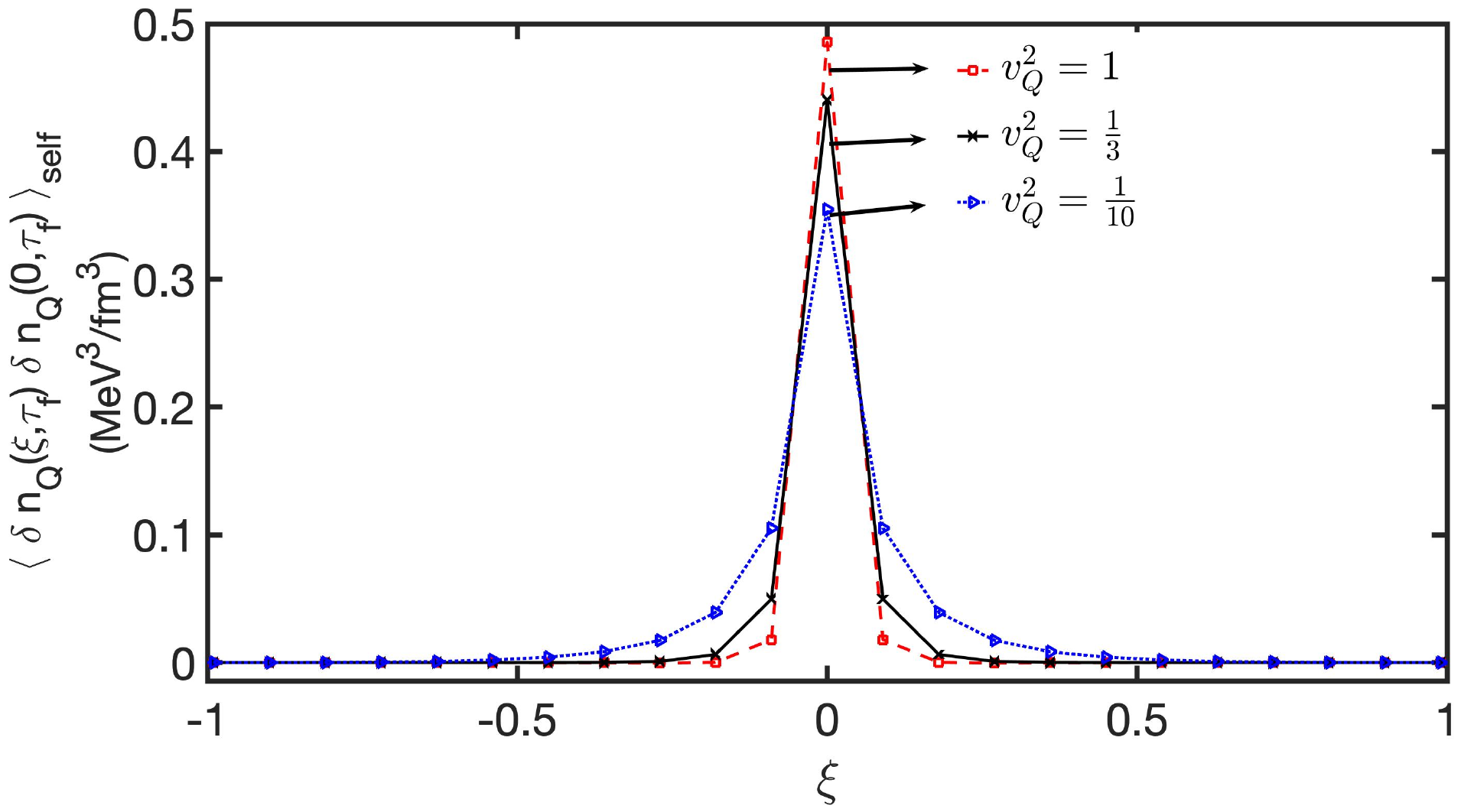}
\caption{(Color online) Numerical results for self-correlations for colored noise with different signal propagation speeds.}
\label{fig:1DselfCorr}
\end{figure}

The nature of the self-correlation is easiest to understand in the 1+1 D case. The mathematical expression for self-correlation is given by (Refs. \cite{Kapusta:2017hfi,De:2020yyx}).
\begin{displaymath}
\Big\langle \delta n_Q (\xi_1,\tau_f)\delta n_Q(\xi_2, \tau_f)\Big\rangle_{\text{self}}
 =\frac{\tau_f}{D_Q} \int \frac{dk}{2\pi} e^{i k (\xi_1 - \xi_2)} \nonumber \\
 \end{displaymath}
 \vspace{-15pt}
 \begin{equation}
\times \int \frac{d\tau''}{\tau''}  \; \frac{\tilde{G}(-k,\tau_f,\tau'')}{ik} \; \langle f(\tau_f) f(\tau'')\rangle
\end{equation}
The $\tilde{G}$ is a Green function, see Appendix~\ref{Sec:1DDiffusion}.  Direct numerical simulations in Ref.~\cite{De:2020yyx} showed how the self-correlation looks for various values of $\tau_Q$, as depicted in the Fig.~\ref{fig:1DselfCorr}.
One can see that as $\tau_Q \to 0$, the self-correlation looks more and more like a Dirac $\delta$- function. The absence of the Dirac $\delta$-function in the case of colored noise means that there is no fixed equilibrium value. The definition of self-correlation is straightforwardly extendable to the 3+1 D.

\begin{figure}[ht!]
\centering
\includegraphics[width=0.85\linewidth]{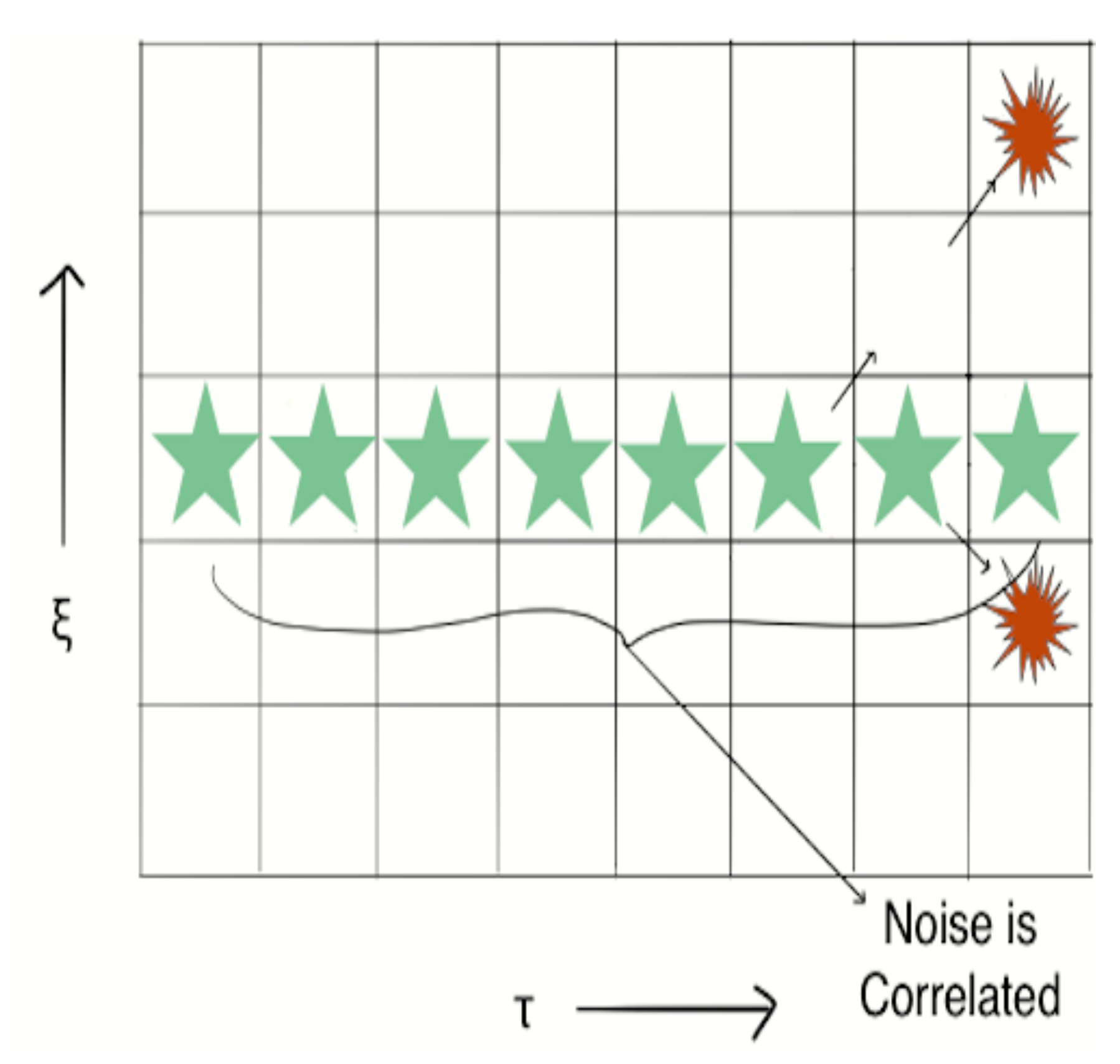}
\caption{(Color online) Description of self-correlation. The stars denote a noise source and the bursts are the charge fluctuations resulting from the noise. Reproduced from Ref.~\cite{De:2020yyx}. }
\label{fig:Schematic_self}
\end{figure}

Self-correlation can be interpreted as the correlation of a charge fluctuation generated in $\xi_1$ at final time $\tau_f$ with another charge fluctuation generated in $\xi_1$ at a previous time but which has traveled to $\xi_2$ at $\tau_f$. It is nontrivial for colored noise because colored noise fluctuations in the same $\xi$ are correlated in time. Figure~\ref{fig:Schematic_self} gives an illustration of self-correlation.

Higher point correlations are also introduced because of causal diffusion. Specifically, the non-linear treatment of the white noise introduces the higher-order correlations.

\newpage

\begin{figure}[h!]
\centering
\includegraphics[scale=0.4]{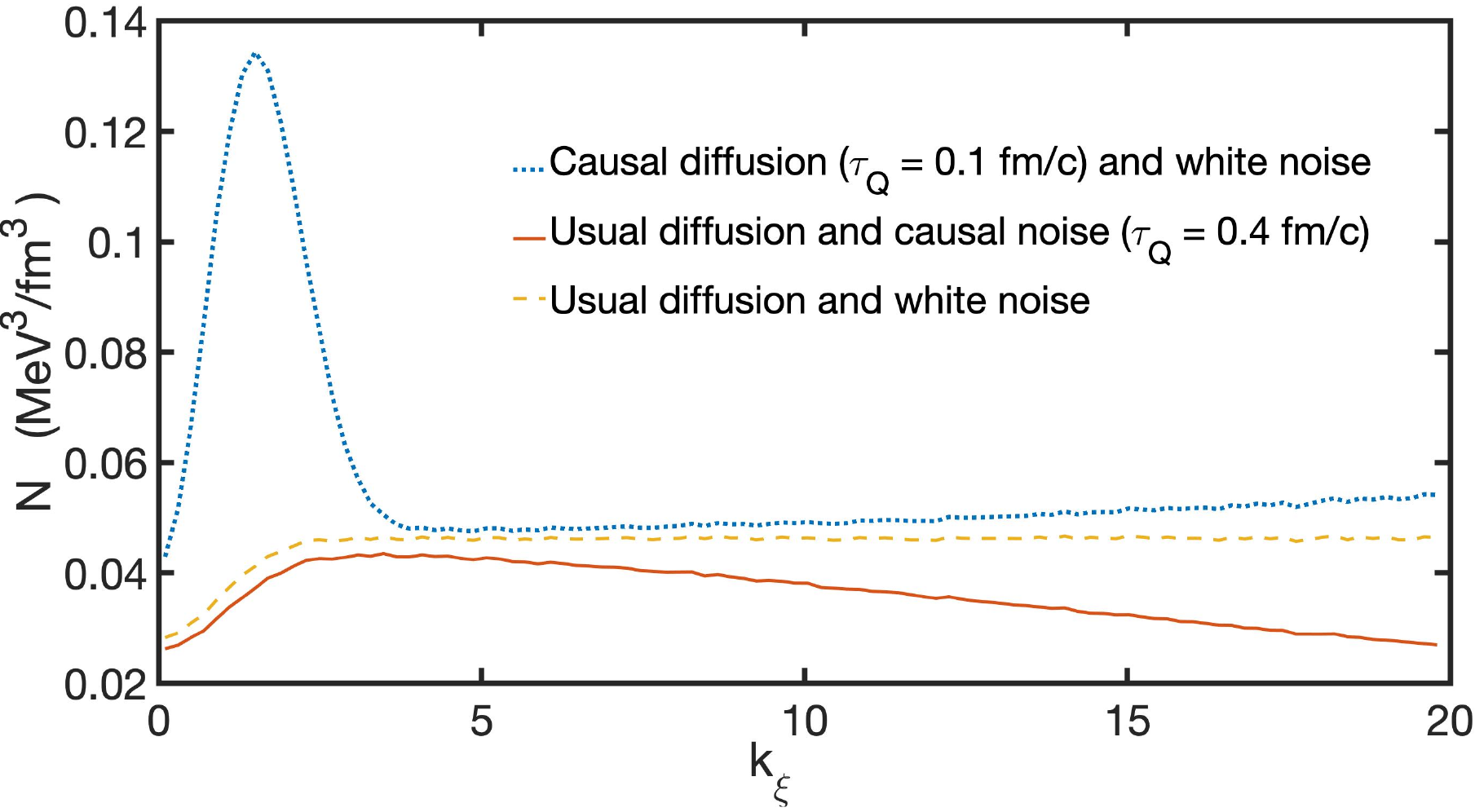}
\caption{(Color Online) The 2-point correlation functions of charge fluctuations in momentum space for three combinations of different types of diffusion equations and noises to disentangle the effects of causal diffusion and causal noise. The plot is computed at $\tau_f = 6.3$ fm/$c$.}
\label{fig:Disentangle_1}
\end{figure}

Given the behavior of the 2-point correlator in the presence of colored noise, with a peak around $k_*$ and a decaying tail, it is interesting to disentangle the effects of causal diffusion and causal noise. Figure~\ref{fig:Disentangle_1} shows the 2-point correlator as a function of $k_{\xi}$ for the following three cases:

\begin{enumerate}
    \item Causal diffusion and white noise.
    \item Usual diffusion and causal noise.
    \item Usual diffusion and white noise 
\end{enumerate}

\begin{figure}[h!]
\centering
\includegraphics[scale=0.4]{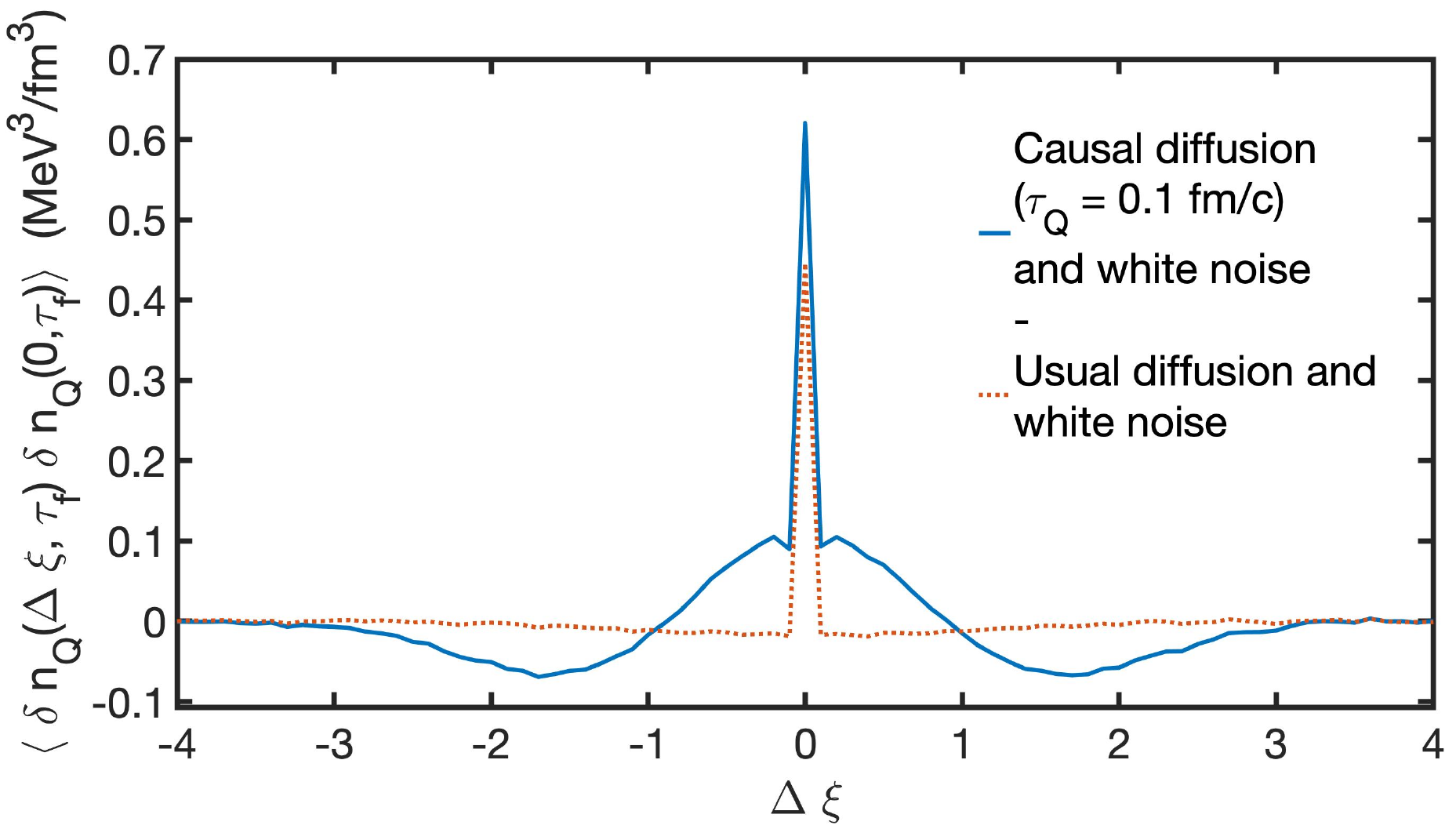}
\caption{(Color online) The effects of causal diffusion evolution with white noise on 2-point charge density correlation in coordinate space with lattice spacing $d\xi =0.1$.}
\label{fig:causalD_whiteN}
\end{figure}

First, consider case 1. Causal diffusion is an inefficient process compared to usual diffusion when it comes to distributing the charge buildup at a particular space point to neighboring points. In addition, white noise fluctuations at a particular point at a specific time are uncorrelated with whatever noise fluctuations that have originated at that point in previous times. This just means that there is continuous pumping of noise fluctuations at a space point through its temporal journey irrespective of the amount of charge buildup there. Causal diffusion does a limited redistribution of the charge buildup. Hence there is a peak in the 2-point correlation around the length scale $k_*$. This is further illustrated in Fig.~\ref{fig:causalD_whiteN} which is a plot of the 2-point correlator in coordinate space. 

\begin{figure}[h!]
\centering
\includegraphics[scale=0.4]{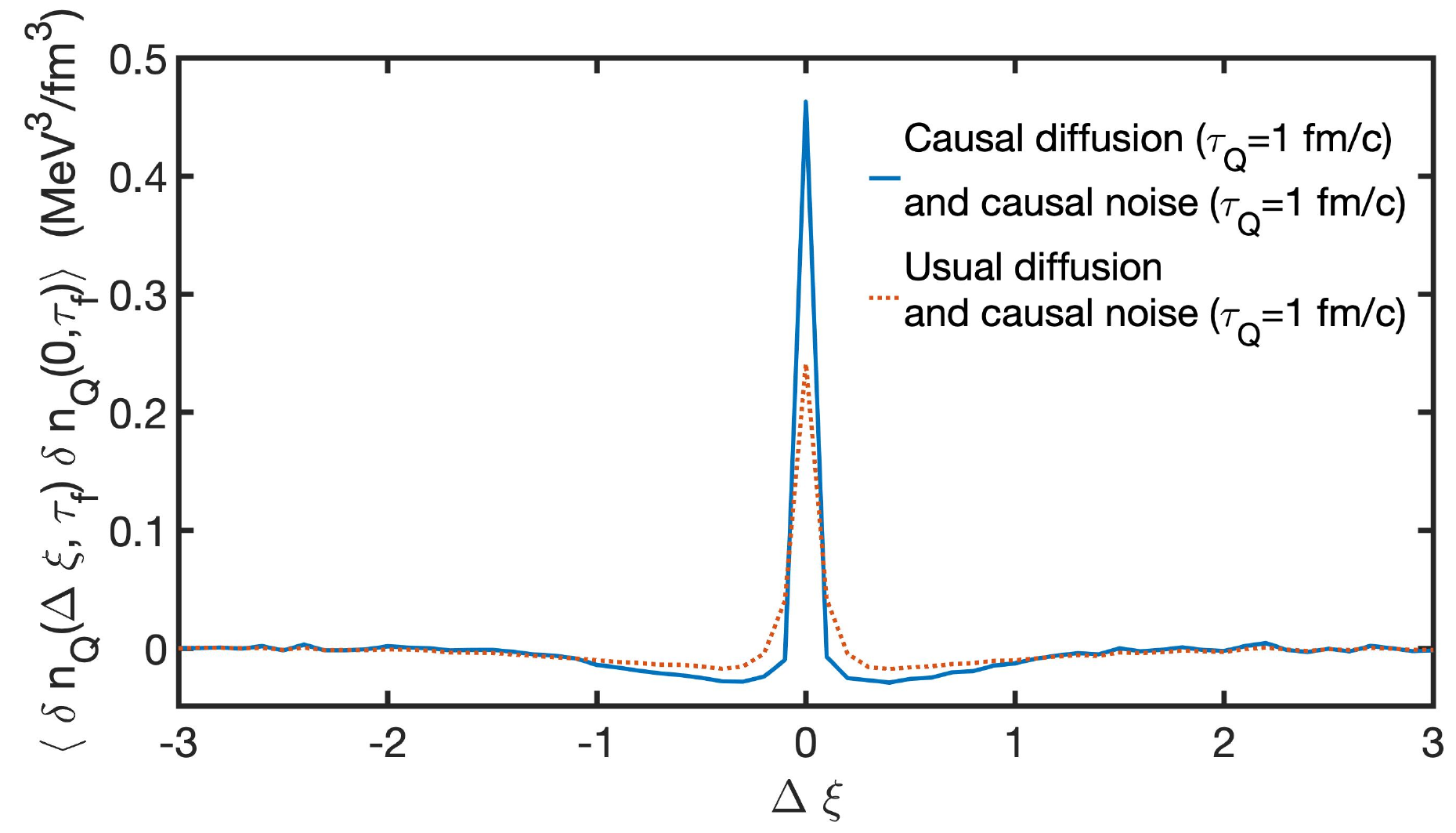}
\caption{(Color online) The effects of causal diffusion evolution with causal noise on 2-point charge density correlation in coordinate space with lattice spacing $d\xi = 0.1$.}
\label{fig:whiteD_causalN}
\end{figure}

Now consider case 2. When we incorporate causal noise, the noise fluctuations that occur at a spatial point are correlated to the noise fluctuations that have already originated there in the previous moments. Hence a spatial point that has had a lot of charge buildup, because of recent large charge fluctuations, will not have large fluctuations at subsequent times because of the exponentially decaying correlation. In this case, the diffusion, either usual or causal, has enough time to distribute the charge buildup effectively. This is further illustrated in Fig.~\ref{fig:whiteD_causalN}.
Both curves in Fig.~\ref{fig:whiteD_causalN} use the same causal noise and hence the self-correlation part for both the curves are the same. The presence of causal diffusion narrows the inverted Gaussian (which signifies the diffusion). Case 3 is the standard white noise result from Fig.~\ref{fig:whiteD_causalN} of Ref.~\cite{De:2020yyx}, specifically, the Dirac $\delta$-function minus a Gaussian. Here, the noise fluctuations pumped in by white noise is effectively dispersed away with infinite signal propagation speed by the usual diffusion process.  

This brings us to our final point. The fluctuation-dissipation theorem tells us that causal noise and causal diffusion are inextricably present. Hence we expect both a peak in the two point correlator and a decaying tail at large $k_{\xi}$. Both may have phenomenological consequences. 

\section{Conclusion}
\label{Sec:Conclusion}

Hydrodynamics is an effective theory that describes the behavior of quark-gluon plasma in a heavy-ion collision. With viscous hydrodynamics comes thermal noise that affect the macroscopic fluid variables. In this paper, we compared the two prevalent ways to study the evolution and effects of thermal fluctuations in relativistic hydrodynamics. The direct stochastic calculation is compared with the deterministic way of tracking the 2-point functions of hydrodynamic variables which are called hydro-kinetics in the literature. Here we considered the relatively simple case of a Bjorken evolution of a single conserved charge, including transverse fluctuations, and focus on the comparison of the 2-point charge correlation functions. We first perform comparisons using white noise and find exact agreements between the two approaches. This agreement is expected because white noise only introduces 2-point correlations. We then extend the stochastic hydrodynamics approach to study the evolution of colored (Catteneo) noise with the causal diffusion equation. The colored noise together with causal evolution results in significant deviations from the results at the white noise limit.

The main result of our work is that the colored noise evolved with the causal diffusion equation generates a peak structure around $k_{\xi} \sim k_*$ and decaying tail at large $k_{\xi}$ in the 2-point correlation function in the Fourier space as shown in Fig.~\ref{fig:3Dcausal1}. These features can be better understood in coordinate space. The peak around $k_*$ is a consequence of the correlation function being confined inside the light cone.
The 2-point charge correlation function at large $k_{\xi}$ is no longer a constant in the causal case because the self-correlation has a finite extension in the Catteneo noise case. Even if the system was non-expanding, the Catteneo noise would have induced this feature. 
These features in the causal charge correlation functions could lead to phenomenological consequences in final-state particle correlations. Specifically, the enhancement and narrowing of the charge balance functions in the presence of colored noise has been discussed in Ref.~\cite{De:2020yyx}.

The quantity $k_*$ is an important emergent length scale in the dynamically evolving system. Modes with $k_{\xi} \gtrsim 0.5 k_*$ remain in equilibrium whereas those with $k_{\xi} \lesssim 0.5 k_*$ are pulled out of equilibrium because of the background expansion of the system. The value of $k_*$ should not change significantly in the Catteneo noise with causal diffusion. This is because the competition between the quenching rate for the fluctuations and the medium expansion rate is not affected a lot by the non-linearity introduced by the colored noise and causal equation of motion. The $k_*$ for the causal case will be a little less than that of the usual diffusion case. Hence, a possible future direction would be to study observables associated with $k_*$. Another interesting avenue would be to explore the emergent scales when a critical point is introduced into the equation of state. This question was investigated in Ref.~\cite{Akamatsu:2018vjr}.

It has been shown in this work that the hydro-kinetic approach is equivalent to solving the white noise SDE. It remains an open question whether the causal evolution and correlated color noise can be implemented in the hydro-kinetic approach. As shown in Fig.~\ref{fig:4ptcorrelation}, the correlated Catteneo noise introduces higher-order correlations. We also find that the causal diffusion evolution introduces higher-order correlations because of its non-linear treatment of noise. It would be interesting to compare the higher-point correlation functions with the results from the hydro-kinetic approach in Ref.~\cite{An:2020vri}.
 
Finally, it is clear from the discussion at the end of Sec.~\ref{Sec:3DComp} and Fig.~\ref{fig:causalD_whiteN} that considering causal diffusion without colored noise leads to a buildup of charge correlation in small $k_\xi$. The location of this buildup depends on grid size, which is unphysical. Causal diffusion and colored noise need to be considered as a package whenever we are considering the effects of stochastic noise on hydrodynamic variables. In the full-fledged causal Israel-Stewart hydrodynamics \cite{Young:2014pka, Murase:2015oie, Singh:2018dpk, Sakai:2020pjw}, stochastic fluctuations are sourced as white noises in the relaxation-type of equation of motion for shear stress tensor to generate the finite time correlations. Such treatment is consistent with the Catteneo noise generation in our case. One can go further and consider the Gurtin-Pipkin \cite{Gurtin} noise which introduces a noise correlation in the spatial direction in addition to the correlation in time. Gurtin-Pipkin noise is in some sense more `physical' and has been dealt with analytically in Ref.~\cite{Kapusta:2014dja}. Exploration with a noise profile with a spatial correlation is left to future work.

\section{Acknowledgments}
The authors thank Derek Teaney and Mayank Singh for their helpful discussions. This work was supported by the U.S. Department of Energy, Office of Science, Office of Nuclear Physics, under DOE Contract Nos. DE-FG02-87ER40328 and DE-SC0021969. C.S. acknowledges a DOE Office of Science Early Career Award.
We acknowledge the Minnesota Supercomputing Institute (MSI) at the University of Minnesota for providing resources that contributed to the research results reported within this paper.

\appendix
\section{Calculations in 1+1 D}\label{Sec:1DDiffusion}
\begin{figure}[h!]
\centering
\includegraphics[width=0.95\linewidth]{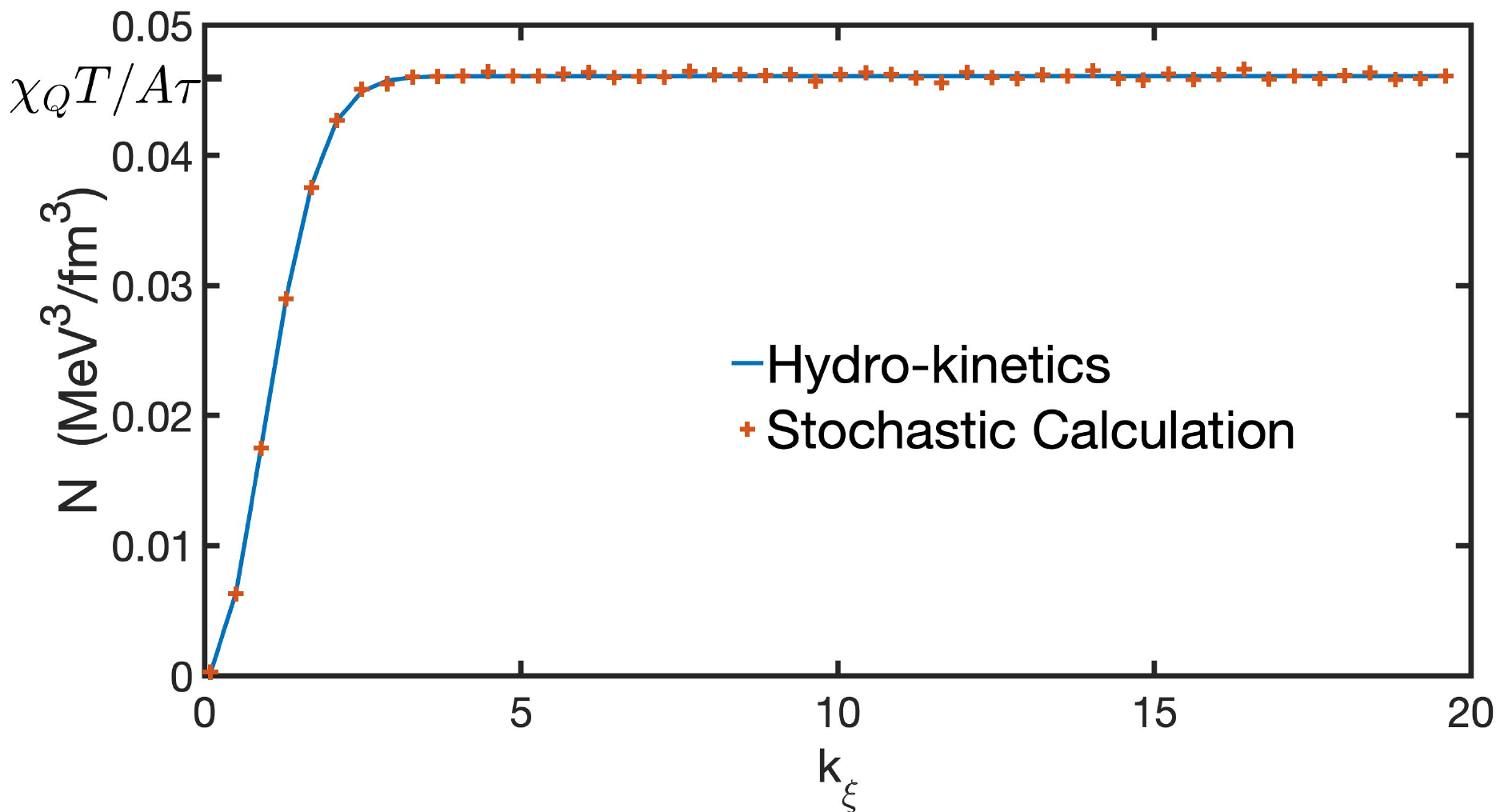}
\caption{(Color online) Comparison of the 2-point charge correlation function in momentum space from the hydro-kinetics and stochastic hydrodynamics with white noise in the 1+1D case and with $A = 1$ fm$^2$.}
\label{fig:1DwhiteNoise}
\end{figure}

It is instructive to understand noise propagation in 1+1 D dynamics.
In this simplified case, start with the white noise diffusion equation given by
\begin{equation} 
\frac{\partial X}{\partial \tau} - \frac{D_Q}{\tau^2}\frac{\partial^2 X}{\partial \xi^2} + s\frac{\partial f}{\partial \xi} = 0 \,.
\label{diffeqwhite}
\end{equation}
The Fourier transform of $X$ is
\begin{equation}
X(\xi,\tau) = \int_{-\infty}^{\infty} \frac{dk}{2\pi } e^{ik\xi} \tilde{X}(k,\tau) \,,
\end{equation}
and similarly for $f$.  Then the SDE for white noise is
\begin{equation}
\frac{\partial }{\partial \tau} \tilde{X} + \frac{D_Q k^2}{\tau^2} \tilde{X} = -iks\tilde{f} \,.
\label{SDEwhite}
\end{equation}
Given a SDE of the form of Eq.~\eqref{SDEwhite} one can write
\begin{equation}
\tau \delta \tilde{n}_Q(k,\tau) = -\int^{\tau_f}_{\tau_i} d\tau' \, s(\tau') \, \tilde{G}(k;\tau,\tau') \, \tilde{f}(k,\tau').
\end{equation}
Then the 2-point correlation function can be written as
\begin{eqnarray}
\langle \tau_1 \delta \tilde{n}_Q(k_1,\tau_1) \tau_2 \delta \tilde{n}_Q(k_2,\tau_2)\rangle 
= \int d\tau'_1 s(\tau'_1 ) \int \tau'_2 s(\tau'_2 ) \qquad \nonumber \\ 
\times \tilde{G}(k_1;\tau_1,\tau'_1) \tilde{G}(k_2;\tau_2,\tau'_2) \langle \tilde{f}(k_1,\tau'_1)\tilde{f}(k_2,\tau'_2)\rangle. \qquad
\label{green_12}
\end{eqnarray}
The Green's function for the homogeneous part of Eq.~\eqref{SDEwhite} is
\be\tilde{G}(k,\tau,\tau') = ik e^{D_Q k^2(\frac{1}{\tau} - \frac{1}{\tau'})}.
\ee
This Green's function is for acausal white noise. 
After evaluating all the integrals one gets
\begin{equation} 
 N(\tau_f,k) =  \frac{\chi_Q T_f}{A \tau_f} 
 \left[ 1 - e^{-2D_Q k^2(\frac{1}{\tau_i}-\frac{1}{\tau_f})} \right]\,.
 \label{exact}
\end{equation}
The large $k$ limit is $\chi_Q T_f/A \tau_f$. 
We compare this to the results of the SDE in Fig.~\ref{fig:1DwhiteNoise}. 
One can see from the figure that the agreement is excellent.  This is expected because white noise only contains 2-point functions and the hydro-kinetic equation just tracks the effect of it. 

Balancing the rate of dissipation to the rate of Bjorken expansion leads to
\begin{equation}
D_Q k_*^2 = \tau_f.
\end{equation}
For the parameter values used in the body of the paper, $k_* \approx 6.26$. We can see from the plot that beyond $k_*$ the $k$-modes asymptote to the value of $\chi_Q T/A \tau$.

The charge conservation equation to be solved for Catteneo noise is \cite{Kapusta:2017hfi}
\begin{displaymath}
\frac{\tau_Q \tau }{D_Q\chi_Q T} \frac{\partial^2 \tilde{X}}{\partial \tau^2} +
\left[ \frac{\tau}{D_Q\chi_Q T} + \tau_Q \frac{\partial}{\partial \tau} \left(\frac{\tau}{D_Q\chi_Q T}\right) \right] \frac{\partial \tilde{X}}{\partial \tau}.
\end{displaymath}
\vspace{-15pt}
\begin{displaymath}
+ \frac{k^2}{\tau \chi_Q T} \tilde{X}
= -i \frac{k \tau_Q \tau s}{D_Q\chi_Q T}\frac{\partial \tilde{f}}{\partial \tau}
\end{displaymath}
\vspace{-15pt}
\be
-ik \left[ \frac{\tau s}{D_Q\chi_Q T}+ \tau_Q \frac{\partial}{\partial \tau}\left(\frac{\tau s}{D_Q\chi_Q T}\right)\right] \tilde{f}.
\label{color-kspace}
\ee
In Ref. \cite{Kapusta:2017hfi} it was assumed that $\tau_Q$ and $D_Q$ were constant while $\chi_Q$ was proportional to $T^2$ and $s$ was proportional to $T^3$.  That allowed for the Green's function to be calculated and expressed in terms of special functions.
One can use Eq.~\eqref{green_12} to compute the 2-point correlator in k-space. Alternatively, one can simulate the SDE Eq.~\eqref{color-kspace} numerically and obtain the 2-point correlator that way. The procedure for simulating such stochastic differential equations was given in Ref.~\cite{De:2020yyx}. A brief summary of the numerical stochastic machinery is provided in Appendix~\ref{Sec:stochastic_machinery}. 

\begin{figure}[h!]
\centering
\includegraphics[width=0.95\linewidth]{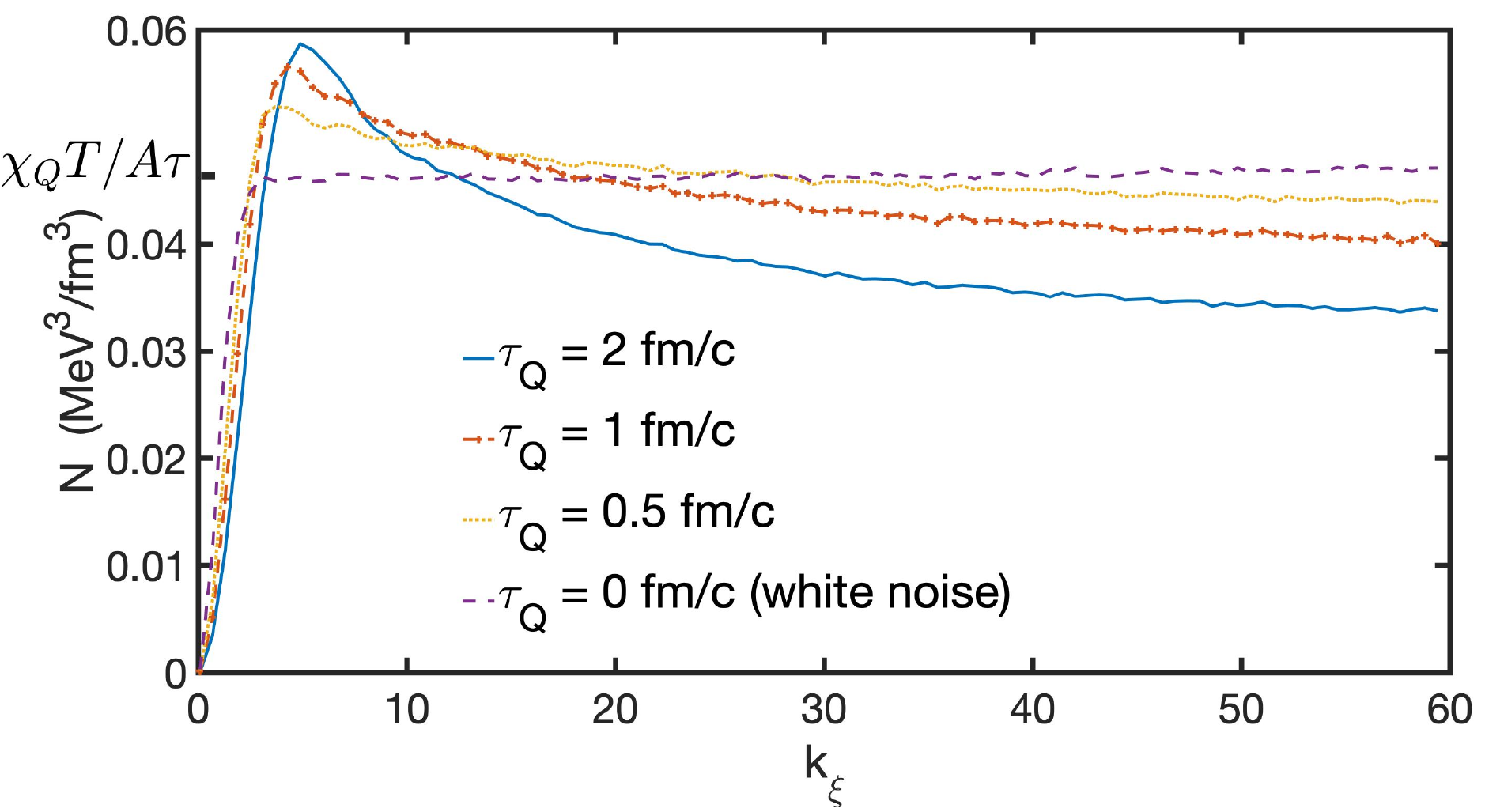}
\caption{(Color online) The 2-point charge correlation function in momentum space with several values of $\tau_Q$ in the (1+1)D system. The plot is at final time $\tau_f = 6.3$ fm/$c$ with $A = 1$ fm$^2$.}
\label{fig:1DcausalNoise}
\end{figure}

\begin{figure}[t]
\centering
\includegraphics[width=0.85\linewidth]{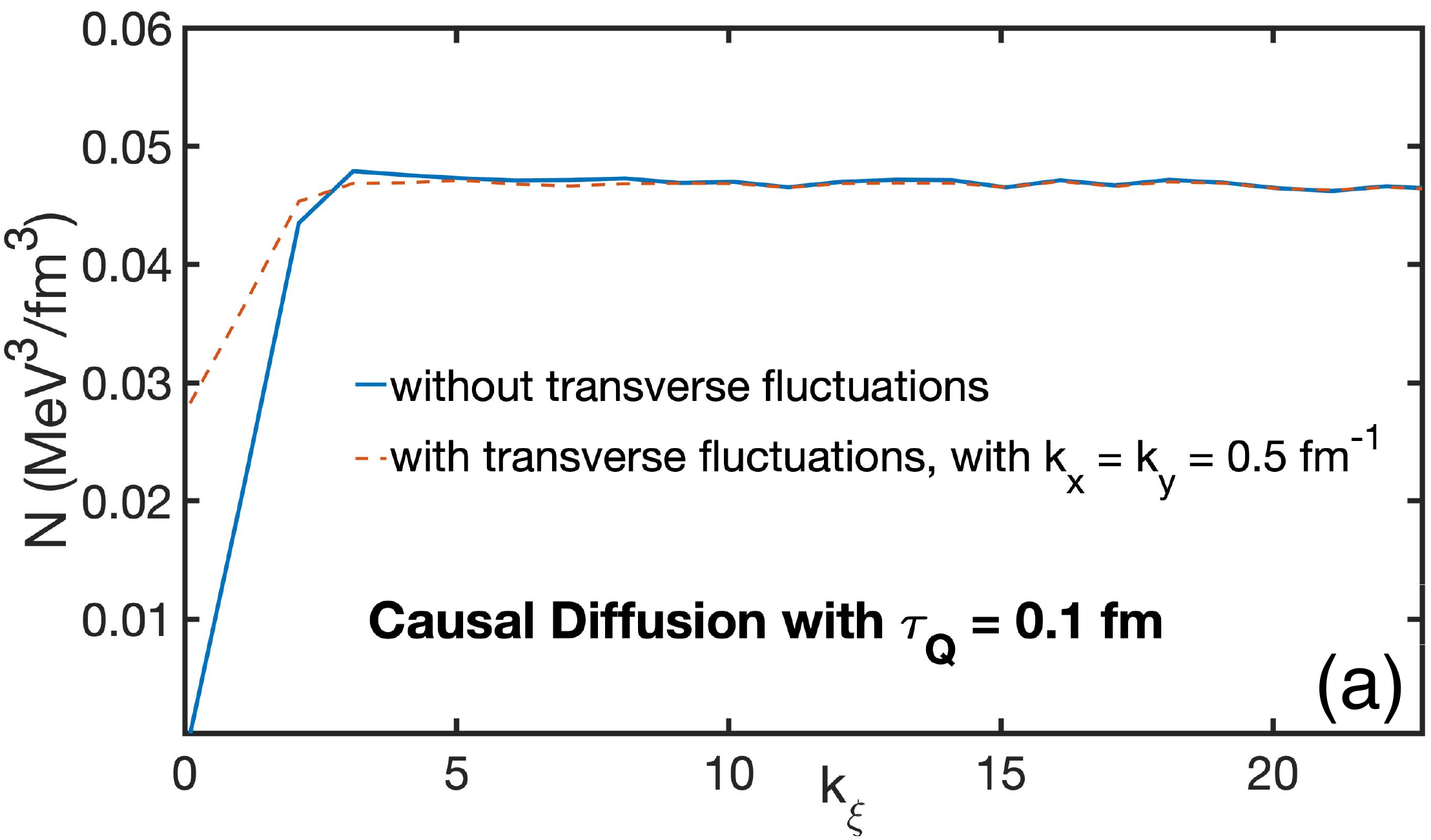}
\includegraphics[width=0.85\linewidth]{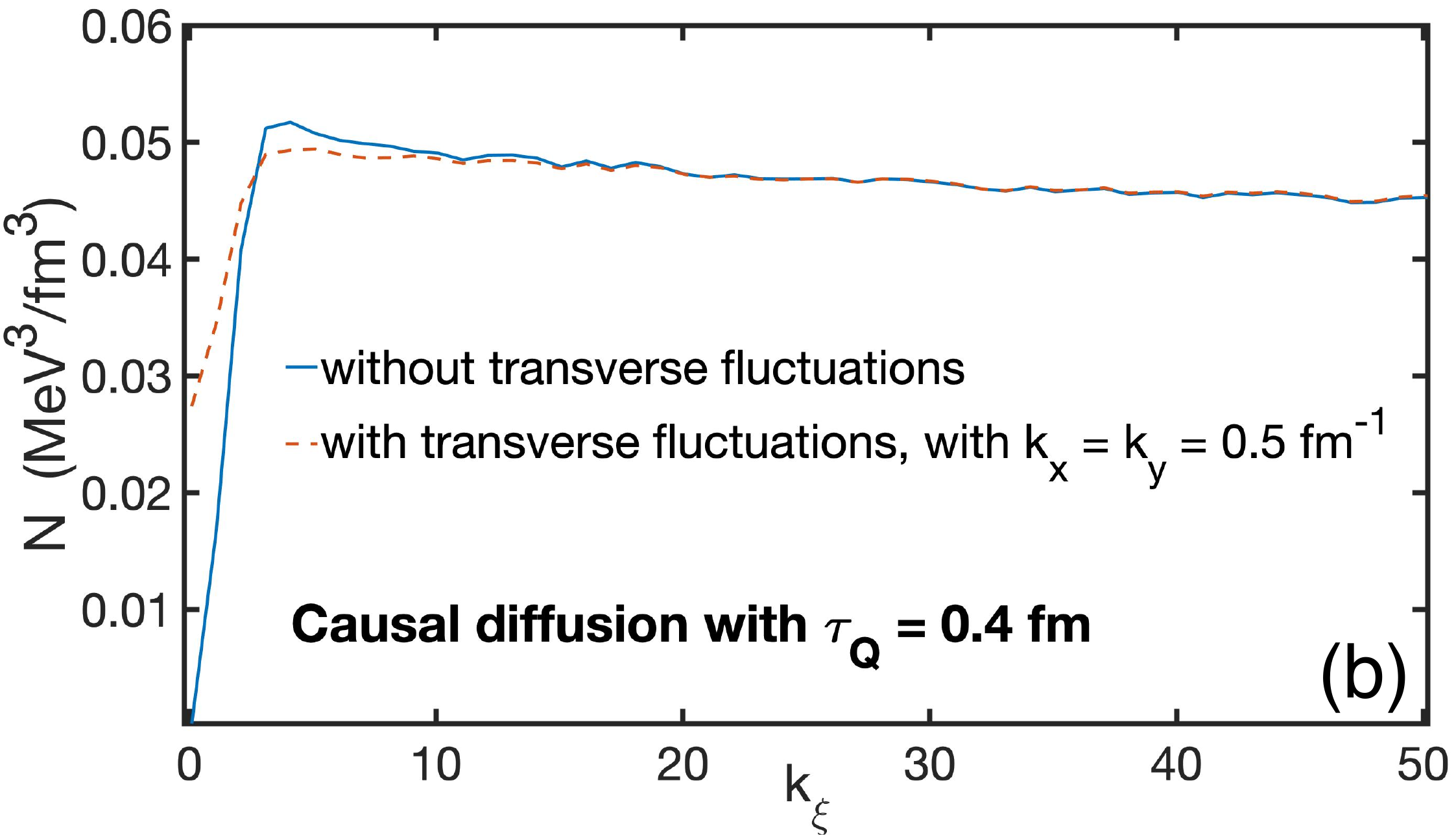}
\caption{(Color online) Effects of transverse fluctuations in the presence of causal Catteneo noise on the 2-point charge correlation functions. (a) is for $\tau_Q = 0.1$\,fm/$c$ and (b) is for $\tau_Q = 0.4$\,fm/$c$. Both sets are computed with $10^5$ events.}
\label{fig:transverse_comparison}
\end{figure}

A plot of $N$ for Catteneo noise for various values of $\tau_Q$ is shown in Fig.~\ref{fig:1DcausalNoise}.  Additionally, we see that in the presence of transverse fluctuations, even though the 2-point correlator for the small $k$ values differs, the large $k$ values for the 2-point correlation for charge fluctuation are the same with or without transverse diffusion. This is illustrated in Fig.~\ref{fig:transverse_comparison}.

\section{Stochastic calculus on a discrete lattice}\label{Sec:stochastic_machinery}

The 2-point correlation of a function is given by
\begin{equation}
 \langle f(x)f(x') \rangle  = \displaystyle{\frac{\displaystyle{\sum_{\mathrm{all \: random \: events}}} f(x) f(x') \quad\quad  }{\mathrm{number \: of \: random \: events}}}.
\end{equation}
In the continuum limit a white noise random function is defined as
\begin{equation}
 \langle f(x)f(x') \rangle = M(x) \delta(x-x') \text{ and } \langle f(x) \rangle = 0
\label{whiteDef}
\end{equation}
where $M(x)$ is a normalization factor and all higher-order cummulants are required to vanish.  The event-by-event distribution of $f$'s fluctuations is therefore a normal distribution with finite variance and zero mean. One can infer that, after discretization, we will have
\begin{equation} 
\langle f(x_i)f(x_{i'})\rangle = \frac{\delta_{ii'}}{\Delta x}.
\end{equation}
The normalization factor in the above equation has been set to unity. The $\delta_{ii'}/\Delta x$ becomes a Dirac delta function in the continuum limit $\Delta x \rightarrow 0$. Therefore, we sample the white noise function $f$ from a normal distribution of mean $0$ and standard deviation $1/\sqrt{\Delta x}$.
We use a random number generator for a large number of instances (e.g. $10^7 $) to minimize statistical noise. Subsequently, one finds the following relations for correlation functions on a discrete lattice

\begin{eqnarray}
\langle f(x_i)f'(x_{i'})\rangle = \frac{\delta_{i+1,i'}-\delta_{i,i'}}{\Delta x^2} \hspace{1.7cm} \end{eqnarray}

\begin{eqnarray}
\langle f'(x_i)f'(x_{i'})\rangle = -\frac{\delta_{i,i'+1}+\delta_{i,i'-1}-2\delta_{i,i'}}{\Delta x^3} 
\end{eqnarray}

\begin{equation}
\left\langle \int_{x_i}^{x_1} f(x') dx' \int_{x_i}^{x_2} f(x') dx' \right\rangle = \min(x_1,x_2) - x_i
\end{equation}

\bibliography{hydrokinetics}

\end{document}